\documentclass[12pt]{article}
\usepackage{latexsym}
\usepackage{amsmath}
\usepackage{amsfonts}
\usepackage{amssymb}
\usepackage{graphicx}
\usepackage{color}
\begin{document}
\input epsf
\def\be{\begin{equation}}
\def\bea{\begin{eqnarray}}
\def\ee{\end{equation}}
\def\eea{\end{eqnarray}}
\def\d{\partial}
\definecolor{red}{rgb}{1,0,0}
\long\def\symbolfootnote[#1]#2{\begingroup%
\def\thefootnote{\fnsymbol{footnote}}\footnote[#1]{#2}\endgroup}
\renewcommand{\a}{\left( 1- \frac{2M}{r} \right)}
\newcommand{\dm}{\begin{displaymath}}
\newcommand{\edm}{\end{displaymath}}
\newcommand{\com}[2]{\ensuremath{\left[ #1,#2\right]}}
\newcommand{\la}{\lambda}
\newcommand{\eps}{\ensuremath{\epsilon}}
\newcommand{\half}{\frac{1}{2}}
\newcommand{\field}[1]{\ensuremath{\mathbb{#1}}}
\renewcommand{\l}{\ell}
\newcommand{\bl}{\left(\l\,\right)}
\newcommand{\normljk}{\langle\l,j,k|\l,j,k\rangle}
\newcommand{\N}{\mathcal{N}}
\renewcommand{\b}[1]{\mathbf{#1}}
\renewcommand{\v}{\xi}
\newcommand{\tr}{\tilde{r}}
\newcommand{\ttheta}{\tilde{\theta}}
\newcommand{\tgamma}{\tilde{\gamma}}
\newcommand{\bg}{\bar{g}}
\renewcommand{\implies}{\Rightarrow}
\newcommand{\z}{\ensuremath{\ell_{0}}}
\newcommand{\temp}{\ensuremath{\sqrt{\frac{2\z+1}{\z}}}}
\newcommand{\twomatrix}[4]{\ensuremath{\left(\begin{array}{cc} #1 & #2
\\ #3 & #4 \end{array}\right) }}
\newcommand{\columnvec}[2]{\ensuremath{\left(\begin{array}{c} #1 \\ #2
\end{array}\right) }}
\newcommand{\e}{\mbox{\textbf{e}}}
\newcommand{\gm}{\Gamma}
\newcommand{\bt}{\bar{t}}
\newcommand{\bphi}{\bar{\phi}}
\newcommand{\m}{\ensuremath{\mathbf{m}}}
\newcommand{\n}{\ensuremath{\mathbf{n}}}
\renewcommand{\theequation}{\arabic{section}.\arabic{equation}}
\newcommand{\newsection}[1]{\section{#1} \setcounter{equation}{0}}
\newcommand{\p}{p}
\newcommand{\tmu}{\tilde{\mu}}
\newcommand{\slthree}{\mathrm{SL}(3,\mathbb{R})}
\newcommand{\sech}{\,\mbox{sech}\,}
\newcommand{\csch}{\,\mbox{csch}\,}
\newcommand{\Pc}{\mathcal{P}}
\newcommand{\Qc}{\mathcal{Q}}
\newcommand{\Jc}{\mathcal{J}}
\newcommand{\Mc}{\mathcal{M}}

\vspace{20mm}
\begin{center} {\LARGE Non-supersymmetric microstates of the D1-D5-KK system  }
\\
\vspace{20mm} {\bf  Stefano Giusto$^{1}$, Simon F. Ross$^{2}$ and Ashish Saxena$^{1}$}\\ 
\vspace{2mm}
\symbolfootnote[0]{ {\tt giusto@physics.utoronto.ca, S.F.Ross@durham.ac.uk} and {\tt ashish@physics.utoronto.ca}} 
$^{1}$Department of Physics,\\ University of Toronto,\\ 
Toronto, Ontario, Canada M5S 1A7;\\
\vspace{2mm}
and\\
\vspace{2mm}
$^{2}$Centre for Particle Theory,\\ Department of Mathematical Sciences,\\
University of Durham,\\ South Road, Durham DH1 3LE, U.K.\\
\vspace{4mm}
\end{center}
\vspace{10mm}
\begin{abstract}
  We construct a discrete family of smooth non-supersymmetric three
  charge geometries carrying D1 brane, D5 brane and Kaluza-Klein
  monopole charges in Type IIB supergravity compactified on a
  six-torus, which can be interpreted as the geometric description of
  some special states of the brane system. These solutions are
  asymptotically flat in four dimensions, and generalise previous
  supersymmetric solutions. The solutions have a qualitatively similar
  structure to previous non-supersymmetric smooth solutions carrying
  D1 and D5 brane charges in five dimensions, and indeed can be viewed
  as the five-dimensional system placed at the core of a Kaluza-Klein
  monopole. The geometries are smooth,  free of horizons and do not have closed
  timelike curves. One notable difference from the five-dimensional
  case is that the four-dimensional geometry has no ergoregion.
\end{abstract}

\thispagestyle{empty}
\newpage
\setcounter{page}{1}

\newsection{Introduction}

An important goal for the study of black hole thermodynamics is to
understand the gravitational description of the microstates
responsible for the entropy. String theory and in particular the
AdS/CFT correspondence offers the tools needed to explore these
issues. The past few years have seen significant progress in our
understanding of the geometrical description of the states underlying
some special black holes which can be embedded in string
theory. Though a generic microstate responsible for the black hole
entropy is expected to admit a description only in the full string
theory, there is at least a subset of these states which can be well
described by supergravity solutions. Probably the best studied example
is the supersymmetric black hole in five
dimensions~\cite{sen,bmpv}. The microstates of the five dimensional
black hole with two charges, which has a string-scale horizon if we
take into account higher derivative corrections, have been completely
described~\cite{bal}-\cite{skend3}. For the black hole with three
charges in five dimensions, which has a macroscopic horizon, many
explicit examples of the microstates are
known~\cite{mss}-\cite{bena4}, though the picture is far less
complete.  Similar results have been achieved for the case of three
and four charge systems in four
dimensions~\cite{benakraus}-\cite{bgl}. For a review of some of these
developments, see~\cite{mathurreview,benareview}.

The results mentioned above refer to systems with unbroken
supersymmetry in four or five dimensions. It is an important and
non-trivial task to extend the success of the supersymmetric case to
the more general non-supersymmetric states. Although the
supersymmetric black holes already have finite horizon areas, the
non-supersymmetric ones are qualitatively different: notably, because
they have a non-zero temperature. This implies that the study of
non-supersymmetric black holes is significantly more complex; it will
involve issues like Hawking radiation and dynamical
instabilities. Also, from a technical point of view, the task of
finding supersymmetric microstates is greatly facilitated by the
classification theorems in supergravity which hold in the presence of
some unbroken supersymmetry~\cite{sugrasol}. For the
non-supersymmetric case, these techniques are not available.

The only known geometries describing non-supersymmetric microstates
are the ones of~\cite{nss5d}. In the Type IIB duality frame, these
solutions carry D1, D5 and momentum charges in five dimensions. A
natural problem is to extend these solutions by adding a Kaluza-Klein
(KK) monopole charge to the system, to produce non-supersymmetric
microstates of the four-charge system in four dimensions. In the
supersymmetric case, the analogous problem can be solved in a
systematic manner. The results of~\cite{sugrasol} imply that a large
class of supersymmetric solutions can be described by a set of
harmonic functions. In this language, adding KK monopole charge turns
out to be equivalent to adding appropriate constants to some of these
harmonic functions. However, the analysis of~\cite{glr} has shown that
the linear structure underlying the supersymmetric solutions is
completely destroyed when we pass to the non-supersymmetric
case. Thus, the solution of this problem will require the use of
different techniques. We will approach this problem by the same route
taken to construct the five-dimensional non-supersymmetric microstates
in~\cite{nss5d}. We will first construct a suitable general family of
stationary geometries, and then find constraints on the parameters to
obtain smooth solutions.

Qualitatively, we would expect the relevant solutions to look like the
five-dimensional solutions of~\cite{nss5d} placed at the core of a KK
monopole. We could attempt to directly add the KK monopole charge to
the general metric considered in~\cite{nss5d}, which was first
obtained in~\cite{Cvetic}. However, adding the KK monopole charge to
the charged solution would be quite complicated. Instead, we observe
that solutions with D1, D5 and momentum charges can be obtained by
starting from a suitable vacuum solution and applying a sequence of
boosts and dualities. For the solution of~\cite{Cvetic}, the relevant
vacuum solution was the Myers-Perry black hole. We can add the KK
monopole charge to this vacuum ``seed'' solution, and then
subsequently add the other charges. This is a useful way to proceed
because there are powerful solution-generating transformations for the
vacuum solutions, based on an $\slthree$ symmetry of the equations of
motion~\cite{maison}. This solution-generating transformation was used
to construct black hole solutions with KK electric and magnetic
charges in~\cite{rasheed,larsen}. It has recently been shown that it
can be used to add KK monopole charge to any stationary, axisymmetric
solution of the vacuum equations~\cite{gs,fgps}. The black hole
solutions of~\cite{rasheed,larsen} might appear at first glance to
provide appropriate ``seeds'' for us, but they correspond only to
under-rotating versions of the Myers-Perry black hole placed at the
core of the KK monopole, while smooth solutions are obtained by
considering over-rotating black holes. In section \ref{vac}, we
therefore construct new seeds, starting from the Kerr-Bolt
instanton. Once we add the KK monopole charge, the solutions we obtain
will turn out to be an analytic continuation (in parameter space) of
the solutions of~\cite{rasheed,larsen}, and they indeed describe an
over-rotating Myers-Perry black hole at the core of the KK
monopole. The general solution carries KK electric and magnetic
charges and angular momentum in four dimensions.  The KK electric
charge and angular momentum in four dimensions correspond to the two
independent angular momenta in five dimensions, so we would expect
them to be determined in terms of the other conserved charges when we
obtain a smooth solution.

Once we have obtained appropriate vacuum ``seed'' solutions, we add D1
and D5 charges by a sequence of boosts and dualities in section
\ref{sectcharges}.  In this paper we restrict the analysis to the case
with zero momentum charge. The general case has some additional
complications which will be studied in a forthcoming publication. The
solution is given in section \ref{summary}; the reader not interested
in the details of its construction can skip to this point.

In section \ref{sectsmooth}, we identify solutions corresponding to
microstates of the brane system by a systematic search of the
parameter space for values at which all the singularities can be
removed. We find that as expected, the smooth solutions are determined
by the D1, D5 and KK monopole charges, and an integer $n\ge1$. For all
values of $n$ greater than $1$ the solutions are non-supersymmetric;
for $n=1$ the solution reduces to the supersymmetric D1-D5-KK
microstate found in~\cite{benakraus}. In section \ref{reg}, we verify
that the solutions identified in section \ref{sectsmooth} are free of
horizons, curvature singularities and closed time-like curves, and
that the matter fields are also regular.

In section \ref{sectprops}, we study some properties of the solitons. We
find that there is a limit in which the solutions have a near-core
geometry which is an orbifold of $AdS_{3}\times S^3$; as
in~\cite{nss5d}, obtaining this limit requires a suitable scaling of
the charges. Thus these solutions are good candidates to describe
microstates of the D1-D5-KK black hole. A rather surprising feature of
these solutions is that in the four-dimensional metric, there is no
ergoregion. This is in contrast to the five-dimensional solutions
of~\cite{nss5d}, where all the non-supersymmetric solutions had an
ergoregion. This implies that the instability identified for the
five-dimensional solutions in~\cite{Cardoso} will not appear for these
four-dimensional solutions. Investigating their stability is an
important open problem. We also show that if we write the
four-dimensional solutions as a fibration over a three-dimensional
base space, this base space is identical to that obtained for the
five-dimensional solutions of~\cite{nss5d} in~\cite{glr}. Hence, as
argued in~\cite{glr}, the picture of four-dimensional solutions as built
up out of half-BPS ``atoms'' of~\cite{bgl} does not apply to these
non-supersymmetric solutions. 

In the future, we would like to extend this class of solutions by
adding momentum charge, thus producing non-supersymmetric microstates
of the four charge black hole. This is not as straightforward as one
might imagine, because the three charge solutions constructed here
also carry an induced KK monopole charge along the $y$
direction. Adding momentum along $y$ by boosting in that direction
will therefore produces NUT charges in the solution, which makes it
asymptotically not flat (in four dimensions). It might be possible to
cancel this NUT charge by starting with a seed solution which already
carries some NUT charge. Then one can attempt to cancel the induced
NUT charge against the one present in the seed metric. The details of
the construction, however, are likely to be complicated.

Another important issue to address is the stability of these
solitons. It was shown in~\cite{Cardoso} that the five dimensional
non-supersymmetric microstates of~\cite{nss5d} suffer from a classical
instability which arises from the presence of an ergoregion. We have
shown that the four dimensional geometries we construct here do not
have a four-dimensional ergoregion, so we expect that they do not
suffer from this particular type of instability. It would be very
interesting to investigate other possible instabilities of this
system.

It would also be interesting to relate the geometric picture of the
microstates found here to a microscopic description. It would be
particularly interesting to consider the behaviour of these
microstates as we vary the coupling, along the lines of~\cite{bgl},
and see if they can be related to some quiver gauge theory description
at weak coupling.

\newsection{Over-rotating vacuum solution}
\label{vac}

We begin by constructing a suitable vacuum solution carrying KK
electric and magnetic charges. As explained in the introduction, it is
easier to add the KK monopole charge to the vacuum solution and then
add the D1 and D5 charges, because we can add KK monopole charge to
any five dimensional stationary axisymmetric vacuum solution of
Einstein equation by an $\slthree$ solution-generating
transformation~\cite{gs,fgps}. The resulting general solution will
also carry a KK electric charge; this can be thought of as
corresponding to angular momentum along the fiber direction in the
five-dimensional geometry. We need to construct new vacuum solutions
because the known black hole solutions of~\cite{rasheed,larsen} only
describe under-rotating black holes. On the other hand, in order to construct microstates
one needs a family of solutions containing horizon-free geometries.  We could construct appropriate solutions by applying the
procedure of~\cite{gs,fgps} to the over-rotating Myers-Perry
solution. One finds, however, that these solutions lie in the same $\slthree$-orbit as the Kerr-Bolt
instanton trivially lifted to five dimensions. Hence one can equivalently construct the required
vacuum solution by applying an SO(2,1) transformation to the Kerr-Bolt instanton. This 
construction has the advantage of providing a parametrization which is similar to the one
used in~\cite{rasheed, larsen} and, in fact, the solution we obtain is related to 
the one of~ \cite{rasheed, larsen} by a simple analytic
continuation in parameter space.

\subsection{The solution generating technique}
Let us briefly review the solution generating technique of
\cite{maison}. A stationary solution of five-dimensional Einstein
equations can be brought to the form \be ds^2_5 = g_{ab}
(d\xi^a+{\omega^a}_i dx^i)(d\xi^b+{\omega^b}_j dx^j)+{1\over \tau}
ds^2_3\,,\quad \tau=-\mathrm{det}g_{ab} ,
\label{dec}
\ee 
where $a,b=0,1$ and $\xi^0\equiv t$, $\xi^1\equiv z$. $z$ is a compact coordinate and ${\partial\over \partial z}$ is assumed to be
Killing. $\omega^a$ are gauge fields on the three-dimensional space
parametrized by $x^i$, and thus they can be dualized to scalars,
$V_a$, such that
\be
d V_a = -\tau g_{ab} *_3 d\omega^a ,
\ee
where $*_3$ is performed with the metric $ds^2_3$. Introduce the $3\times 3$ unimodular matrix
\be
\chi=\begin{pmatrix}g_{ab} -{1\over \tau} V_a V_b & {1\over \tau} V_a\cr {1\over \tau} V_b & -{1\over \tau}  \end{pmatrix} .
\ee
The equations of motion can be written as
\be
d *_3(\chi^{-1} d\chi)=0
\label{eqone}
\ee
and
\be
R^{(3)}_{ij}={1\over 4}\mathrm{Tr} (\chi^{-1}\partial_i \chi\,\chi^{-1}\partial_j \chi) .
\ee
As shown in~\cite{gs}, it is useful to interpret Eq. (\ref{eqone}) as the integrability condition for the following:
\be
\chi^{-1} d\chi = *_3 d\kappa . 
\ee
This defines a $3\times 3$ matrix of 1-forms $\kappa$. One has that
\be
\omega^0 = -{\kappa^0}_2\,,\quad \omega^1=-{\kappa^1}_2 .
\ee
The equations of motion are invariant under the linear transformation
\be
\chi\to  N\chi N^T\,\,,\,\, \kappa\to (N^T)^{-1} \kappa N^T\,,\quad N\in \mathrm{SL}(3,\mathbb{R})
\label{transformation}
\ee
if the base metric $ds^2_3$ is kept fixed. This $\slthree$ group of transformations can be used to generate new solutions from
known ones. If one wants to preserve the asymptotic structure of the solution, which in our case is $\mathbb{R}^{3,1}\times S^1$, 
the transformation matrix $N$ has to be restricted to the subgroup  $\mathrm{SO}(2,1)$.

\subsection{Constructing the vacuum seed metric}

We want to construct a vacuum solution with the following properties: it goes asymptotically to  
$\mathbb{R}^{3,1}\times S^1$, it carries KK electric and magnetic charges along the $S^1$ and, when the size of the 
KK monopole is made much larger than any other length scale, the solution reduces to the {\it over-rotating} Myers-Perry 
solution. We will obtain such a solution by applying an $\mathrm{SO}(2,1)$ transformation to the following starting metric:
\bea
ds_5^2 &=& -dt^2 + \frac{\tilde{F}}{\rho^2 - (m - b \cos\theta)^2 }\left(dz -\frac{2 m \tilde{\Delta}(m-b \cos\theta)}{b \tilde{F} } d\phi\right)^2 \nonumber \\
& & + (\rho^2-(m - b\cos\theta)^2)\left[ \frac{d\rho^2}{\tilde{\Delta} }+d\theta^2  + \frac{\tilde{\Delta} }{\tilde{F} } \sin^2\theta d\phi^2 \right] ,
\label{kerr-bolt}
\eea
where $\tilde{F}$ and $\tilde{\Delta}$ are
\be
\tilde{F} = \rho^2 + m^2 - b^2 \cos^2\theta, \ \ \tilde{\Delta} = \rho^2+ m^2 - b^2 .
\ee
This is a Kerr-Bolt instanton lifted to five dimensions by adding a flat time direction.
The $\chi$ and $\kappa$ matrices associated to the metric (\ref{kerr-bolt}) are
\be
\chi =\left(\begin{array}{ccc} -1 & 0 & 0 \\ 0 & \frac{ \rho^2- (m + b \cos\theta)^2}{\tilde{F}} & -\frac{2 m \rho}{\tilde{F}} \\
0 & -\frac{2 m \rho}{\tilde{F}} & - \frac{\rho^2 - (m - b \cos\theta)^2 }{\tilde{F} } \end{array} \right) ,
\ee
\be
\kappa = \left( \begin{array}{ccc}  0 & 0& 0 \\ 0 & \frac{2 m b \rho\sin^2\theta}{\tilde{F} } & \frac{2 m \tilde{\Delta}(m-b \cos\theta)}{b \tilde{F} }-\frac{2m^2}{b} \\ 0& 2m\left( \cos\theta-\frac{b\sin^2\theta (m + b \cos\theta )}{\tilde{F} } \right) & -\frac{2 m b \rho\sin^2\theta}{\tilde{F} } \end{array} \right) d\phi .
\ee
Of particular interest to us is the asymptotic behavior of $\kappa$. This is important in determining the condition for the absence of NUT charge in the solution obtained after a general SO(2,1) rotation. We find that
\be
\kappa \approx \left( \begin{array}{ccc}  0 & 0& 0 \\ 0 & 0  & -2 m \cos\theta \\ 0& 2 m \cos\theta & 0 \end{array} \right)
\ee
for large $\rho$. Under a transformation $N$, $\kappa$ transforms as in (\ref{transformation}); using also the fact that 
$\omega^0 = - {\kappa^0}_2$, we see that the transformed solution is free of NUT charge if the $(0,2)$ component of the transformed
$\kappa$ vanishes at large $\rho$. This leads to the condition
\be
N_{13} N_{32} = N_{12} N_{33} \label{nutcharge} .
\ee
A general $\mathrm{SO}(2,1)$ matrix can be decomposed as $N= N_3 N_2 N_1$ where
\be
N_1 = \left( \begin{array}{ccc}  \cosh\alpha & \sinh\alpha& 0 \\ \sinh\alpha  & \cosh\alpha  & 0 \\ 0& 0 & 1 \end{array} \right) ,
\ee
\be
N_2 = \left( \begin{array}{ccc}  1 & 0 & 0 \\  0 & \cosh\beta  & \sinh\beta \\ 0&\sinh\beta  & \cosh\beta \end{array} \right) ,
\ee
\be
N_3 = \left( \begin{array}{ccc}  \cos\gamma & 0& -\sin\gamma \\ 0 & 1  & 0 \\ \sin\gamma & 0 & \cos\gamma \end{array} \right) .
\ee
Using this parametrization of $N$, we can rewrite the NUT elimination
condition (\ref{nutcharge}) as
\be
\tan 2\gamma = \tanh \alpha \csch \beta. \label{nutcond}
\ee
In order to impose this condition we will find it most convenient to solve the above equation for $\alpha$, leaving $\beta$ and 
$\gamma$ as free parameters. Using now the transformation rule (\ref{transformation}), and reconstructing the components of the
transformed metric from the transformed $\chi$ and $\kappa$, we arrive at the following metric:
\bea
\label{seed}
 ds_{5}^2 &= & \frac{B}{A} \left( dz+ A_\mu dx^{\mu} \right)^2 - \frac{f^2}{B} (dt+ \omega^{0}_{\phi} d\phi)^2 + A\left( \frac{dr^2}{\Delta} + d\theta^2 + \frac{\Delta}{f^2} \sin^2\theta d\phi^2 \right) , \nonumber \\
 \eea
 where
 \be
 \Delta = r^2-2M r +P^2 +Q^2-3\Sigma^2-b^2 ,
 \ee
 \be
 f^2 =  r^2-2 M r - b^2\cos^2\theta +P^2 +Q^2-3\Sigma^2 ,
 \ee
\bea
 A_{\mu}dx^{\mu} \!\!\!&=&\!\!\! \frac{C}{B} dt + \left(\omega^{1}_{\phi} +\frac{C}{B} \omega^{0}_{\phi} \right) d\phi, \\
A\!\!\!&=&\!\!\!  \left(r-\Sigma \right)^2 -\frac{2 P^2 \Sigma}{\Sigma-M } -b^2\cos^2\theta + \frac{2 J P Q \cos\theta}{(M+\Sigma)^2 -Q^2} ,\\
B\!\!\!&=&\!\!\!  \left(r+\Sigma \right)^2 -\frac{2 Q^2 \Sigma}{\Sigma+M } -b^2\cos^2\theta - \frac{2 J P Q \cos\theta}{(M-\Sigma)^2 -P^2} ,\\
 C \!\!\!&=&\!\!\!  2 Q \left( r-\Sigma\right) -\frac{2 P J \cos\theta (M+\Sigma )}{(M-\Sigma)^2 -P^2}, \\
 \omega^0_{\phi}\!\!\!&=&\!\!\! \frac{2 J \sin^2\theta}{f^2} \left[ r- M + \frac{(M^2+3\Sigma^2 -P^2-Q^2)(M+\Sigma)}{(M+\Sigma)^2-Q^2}\right] ,\\
 \omega^{1}_{\phi} \!\!\!&=&\!\!\!  \frac{2P\Delta}{f^2} \cos\theta - \frac{2 Q J \sin^2\theta \left[ r(M-\Sigma )+M\Sigma+3\Sigma^2-P^2-Q^2\right]}{f^2\left[ (M+\Sigma)^2-Q^2\right]} .\\
 \eea
  We have redefined the radial coordinate as
  \be
  \rho=r-M.
  \ee
The constants $M,\Sigma,Q,P$ and $J$ are functions of $m$, $b$, $\beta$ and $\gamma$, given by
 \bea
 M &=& \frac{m \sinh\beta\cosh\beta}{\sqrt{1- \sin^2 2\gamma \cosh^2\beta } } ,\\
 P &=& - \frac{m \cos\gamma \left( 1- 2\cos^2\gamma \cosh^2\beta \right) }{\sqrt{1- \sin^2 2\gamma \cosh^2\beta } }, \\
 Q &=&   \frac{m \sin\gamma \left( 1- 2\sin^2\gamma \cosh^2\beta \right) }{\sqrt{1- \sin^2 2\gamma \cosh^2 \beta } }, \\  
 J &=& -\frac{m b \sin2\gamma}{2 } \left( \frac{1 - \sin^2 2\gamma \cosh^4\beta    }{1- \sin^2 2\gamma \cosh^2 \beta  }\right) ,\\
 \Sigma &=& -\frac{m \cos 2\gamma \sinh\beta\cosh\beta}{\sqrt{1- \sin^2 2\gamma \cosh^2\beta } } .
 \label{rasheedconstraints}
 \eea
 It follows from this that the parameters of the solution satisfy the relations
 \bea
 M^2 + 3 \Sigma^2 -P^2 -Q^2 +m^2 &=& 0 , \\
 \frac{Q^2 }{\Sigma +M }+\frac{P^2 }{\Sigma -M} &=& 2 \Sigma , \\
  \frac{ b^2\left[(M+\Sigma)^2 -Q^2 \right]\left[(M-\Sigma)^2 -P^2 \right] }{P^2 +Q^2 -M^2 -3 \Sigma^2 } &=& J^2 .
 \eea
 
In order to perform the dualities of the next subsection, we will also need the potential $V_0$ associated to the metric (\ref{seed}), together with the components ${\kappa^1}_{0,\phi}$ and ${\kappa^0}_{0,\phi}$ of $\kappa$. They are given by
\be
V_0 = -{2\over A}(J \cos\theta+ P Q),
\ee
\bea
\!\!\!\!\!{\kappa^1}_{0,\phi}\!\!\!&=&\!\!\! {2\over f^2}\Bigl[Q \Delta\cos\theta \nonumber\\
&&+ {J P\over (M-\Sigma)^2 -P^2}((r-M)(M+\Sigma) + P^2 +Q^2 -M^2 -3 \Sigma^2)\sin^2\theta\Bigr],\\
{\kappa^0}_{0,\phi}\!\!\!&=&\!\!\! -{2\over f^2}\Bigl[(M+\Sigma)\Delta \cos\theta + {J Q P\over (M-\Sigma)^2 -P^2}(r-M)\sin^2\theta\Bigr] .
\eea

The metric (\ref{seed}) is analogous to the metric found in
\cite{rasheed}, with the crucial difference that while the metric of
\cite{rasheed} goes over to the under-rotating Myers-Perry solution at
the core of the KK monopole, the metric (\ref{seed}) approaches the
over-rotating Myers-Perry solution in the same limit. As for the
metric of \cite{rasheed}, one can rewrite the solution (\ref{seed}) in
a somewhat more convenient parametrization, analogous to the one found
in \cite{larsen}. In this new form, the parameters $\beta$ and
$\gamma$ are exchanged for parameters $p$ and $q$, defined as
\begin{equation} 
p  = M - \Sigma, \quad q = M+ \Sigma.
\end{equation}
The constraints (\ref{rasheedconstraints}) imply
\begin{equation}
P^2 = \frac{ p (p^2+m^2)}{(p+q)}, \quad Q^2 = \frac{ q
  (q^2+m^2)}{(p+q)},
\end{equation}
\begin{equation}
J^2 = b^2 \frac{pq (pq-m^2)^2}{(p+q)^2 m^2}. 
\end{equation}
We also return to the original radial coordinate,
 $\rho = r - M$. Then the metric functions can be rewritten explicitly
 in terms of this parametrization as
\begin{equation} \label{delta}
\Delta = \rho^2 + m^2 - b^2,
\end{equation}
\begin{equation}
f^2 = \rho^2 + m^2 - b^2 \cos^2 \theta,
\end{equation}
\bea
A &=& f^2 + 2p \left[ \rho + \frac{(pq-m^2)}{(p+q)} + b \frac{
    \sqrt{p^2+m^2}\sqrt{q^2+m^2}}{m (p+q)} \cos \theta \right],  \\
B &=& f^2 + 2q \left[ \rho + \frac{(pq-m^2)}{(p+q)} - b \frac{
    \sqrt{p^2+m^2}\sqrt{q^2+m^2}}{m (p+q)} \cos \theta \right], \\
C &=& 2 \frac{\sqrt{q}}{\sqrt{p+q}} \left[ \sqrt{q^2+m^2}(\rho+p) -
    \frac{q \sqrt{p^2+m^2}}{m} b \cos \theta \right], \\
\omega^0 &=& \frac{2 J \sin^2 \theta}{f^2} \left[ \rho -
  \frac{m^2(p+q)}{(pq-m^2)} \right] d\phi,\\
\omega^1 &=& \frac{2 \sqrt{p}}{\sqrt{p+q}} \frac{1}{f^2} \left[
  \sqrt{p^2+m^2} \Delta \cos \theta - \frac{b \sqrt{q^2 + m^2}}{m}
  (\rho p - m^2) \sin^2 \theta \right] d\phi. \label{omega1}
\eea
The quantities needed for the dualities can be
rewritten in this parametrization as
\begin{equation}
\label{v0}
V_0= - \frac{2}{A} \frac{\sqrt{pq}}{(p+q)} \left[ \frac{b(pq-m^2)}{m}
  \cos \theta + \sqrt{p^2+m^2} \sqrt{q^2+m^2} \right],
\end{equation}
\begin{equation}
\label{k10}
{\kappa^1}_{0,\phi} = \frac{2}{f^2} \frac{\sqrt{q}}{\sqrt{p+q}} \left[
  \sqrt{q^2+m^2} \Delta \cos \theta + \frac{\sqrt{p^2+m^2} b
    (\rho q + m^2)}{m} \sin^2 \theta \right],
\end{equation}
\begin{equation}
\label{k00} 
{\kappa^0}_{0,\phi} = - \frac{2}{f^2} q \left[ \Delta \cos \theta +
  \frac{ b \sqrt{p^2+m^2} \sqrt{q^2+m^2}}{m (p+q)} \rho \sin^2 \theta \right].
\end{equation}
This parametrization manifests the fact that the metric (\ref{seed}) is an analytic continuation 
of the metric in \cite{larsen}. The two metrics are related by
\begin{equation}
p_L = 2p, \quad q_L = 2q,
\end{equation}
\begin{equation}
m_L = -i m, \quad a_L = ib,
\end{equation}
where $p_L$, $q_L$, $m_L$ and $a_L$ are the parameters of \cite{larsen}.

\newsection{Adding charges via dualities}
\label{sectcharges}

In the previous subsection we have constructed a solution of the
five-dimensional vacuum Einstein equations, whose asymptotic limit is
$\mathbb{R}^{3,1}\times S^1$. The only charges carried by this
solution are KK electric and KK magnetic charge along the $S^1$, which
we denote by $P_z$ and $KK_z$, respectively.

We can trivially lift this solution to ten dimensions by adding five
flat compact directions, which we denote by $y$ and $z_1,\ldots,z_4$.
By a sequence of boosts and dualities we can add charges corresponding
to D1 branes wrapped along $y$ and D5 branes along $y,z_1,\ldots,z_4$;
we denote these charges as $D1_y$ and $D5_{y1234}$.  A further boost
along $y$ would add $P_y$ charge, but we do not explicitly perform
this last step in this paper. In this way we generate a non-extremal
solution carrying $P_z$, $KK_z$, $D1_y$ and $D5_{y1234}$ charges. When
augmented with the last $P_y$ charge, this solution represents the
most general non-extremal solution with four non-compact dimensions:
all other solutions are related to this one by dualities.

Let us start by introducing some notation. We rewrite the five-dimensional
vacuum solution in (\ref{seed}) as 
\be
ds^2_5 = -(1-H) (dt+\mathcal{A})^2 + ds^2_4,
\ee
where
\bea
&&(1-H)=-g_{tt}\,,\ \mathcal{A} = \omega^0+{g_{tz}\over g_{tt}}(dz+\omega^1)\,,\ g_{tt}={A f^2 -C^2\over A B}\,,\ g_{tz}={C\over A} ,\nonumber\\
&&ds^2_4=-{\tau\over g_{tt}}(dz+\omega^1)^2+{1\over \tau} ds^2_3 \,,\ \tau={f^2\over A} .
\eea
When lifted to ten dimensions this solution becomes
\be
ds^2_{10} = -(1-H) (dt+\mathcal{A})^2 + ds^2_4+dy^2+ds^2_{T^4}\,,\quad ds^2_{T^4}=\sum_{i=1}^4 dz_i^2 .
\ee

\subsection{Duality chain}

In the following we describe the sequence of boosts and dualities
required to add the desired charges. At each step, the charges of the resulting
solution will be given in parenthesis (for brevity, we will omit the starting $P_z$, $KK_z$ charges, that
are present throughout).  Since this procedure is fairly standard
by now, we will be very schematic. The only computationally
challenging step is the dualization of the RR 6-form into the
corresponding 2-form, so we will give more details of this step. For
brevity, we introduce the notation
\be
s_{1,5}=\sinh\delta_{1,5}\,,\quad c_{1,5}=\cosh\delta_{1,5}\,,\quad H_{1,5}=1+H \sinh^2\delta_{1,5} .
\ee
$B^{(2)}$ denotes the NS-NS B-field and $C^{(p)}$ the p-form RR
field. $\Phi$ is the dilaton. All metrics are in string frame. Our conventions for the normalization of the gauge fields and U-duality rules are as given in Appendix A of ~\cite{mathurmultstring}.

\subsubsection{Boost along $y$ with parameter $\delta_5$ ($P_y$)}

The change of coordinates
\be
t \rightarrow c_5 t + s_5 y\,,\quad  y \rightarrow s_5 t + c_5 y
\ee
produces the metric
\bea
ds_{10}^2&=& H_5 \left[ dy - \frac{c_5 s_5 H}{H_5}(dt+c_5 \mathcal{A})+s_5 \mathcal{A} \right]^2 - \frac{(1-H)}{H_5}(dt+c_5 \mathcal{A})^2\nonumber \\&&+ ds_4^2 + ds^2_{T^4} .
\eea
\subsubsection{T-duality along $y$ ($F1_y$)}
\bea
ds^2_{10}&=&H_5^{-1}dy^2 - \frac{(1-H)}{H_5}(dt+c_5 \mathcal{A})^2+ ds_4^2 + ds^2_{T^4} ,\\
B^{(2)}&=&\left[ - \frac{c_5 s_5 H}{H_5}(dt+c_5 \mathcal{A})+s_5 \mathcal{A} \right]\wedge dy ,\\
e^{2\phi}&=&H_5^{-1}.
\eea
\subsubsection{Boost along $y$ with parameter $\delta_1$ ($F1_y-P_y$)}
The transformation
\be
t \rightarrow c_1 t + s_1 y\,,\quad  y \rightarrow s_1 t + c_1 y
\ee
gives
\bea
ds^2_{10}&=&\frac{H_1}{H_5} \left[ dy -\frac{c_1 s_1 H}{H_1}(dt+c_1 c_5 \mathcal{A}) +s_1 c_5 \mathcal{A}\right]^2  \nonumber\\ &&- \frac{(1-H)}{H_1 H_5} \left[ dt+c_1 c_5 \mathcal{A} \right]^2+ ds_4^2 + ds^2_{T^4} ,\\
B^{(2)}&=&- \frac{c_5 s_5 H}{H_5} \left[ \left( dt + c_1 c_5 \mathcal{A}\right) \wedge \left( dy + s_1 c_5 \mathcal{A}\right) \right] +s_5 \mathcal{A}\wedge \left(c_1 dy - s_1 dt \right) ,\\
e^{2\phi}&=&H_5^{-1} .
\eea
\subsubsection{S-duality ($D1_y-P_y$)}
\bea
ds^2_{10}&=&\frac{H_1}{H_5^{1/2}} \left[ dy -\frac{c_1 s_1 H}{H_1}(dt+c_1 c_5 \mathcal{A}) +s_1 c_5 \mathcal{A}\right]^2  \nonumber\\ &&- \frac{(1-H)}{H_5^{1/2} H_1} \left[ dt+ c_1 c_5 \mathcal{A} \right]^2+ H_5^{1/2}(ds_4^2 + ds^2_{T^4}),\\
C^{(2)}&=&- \frac{c_5 s_5 H}{H_5} \left[ \left( dt + c_1 c_5 \mathcal{A}\right) \wedge \left( dy + s_1 c_5 \mathcal{A}\right) \right] +s_5 \mathcal{A}\wedge \left(c_1 dy - s_1 dt \right),\\
e^{2\phi}&=&H_5.
\eea
\subsubsection{T-duality along $T^4$ ($D5_{y1234}-P_y$)}
\bea
ds^2_{10}\!\!&=&\!\!\frac{H_1}{H_5^{1/2}} \left[ dy -\frac{c_1 s_1 H}{H_1}(dt+c_1 c_5 \mathcal{A}) +s_1 c_5 \mathcal{A}\right]^2  \nonumber\\ &&- \frac{(1-H)}{H_5^{1/2} H_1} \left[ dt+c_1 c_5 \mathcal{A} \right]^2+ H_5^{1/2}ds_4^2 + H_5^{-1/2}ds^2_{T^4} ,\label{met}\\
C^{(6)}\!\!&=&\!\!\left[ -\frac{c_5 s_5 H}{H_5} \left[ \left( dt + c_1 c_5 \mathcal{A}\right) \wedge \left( dy + s_1 c_5 \mathcal{A}\right) \right] +s_5 \mathcal{A}\wedge \left(c_1 dy - s_1 dt \right)\right]\wedge dz_i^4 \label{C6},\\
e^{2\phi}\!\!&=&\!\!H_5^{-1} .
\eea
Note that the type IIB action in our conventions only includes $p$-forms with $p\le 4$. Thus the 6-form gauge field generated in the step above has to be dualized to a 2-form by using the electric-magnetic duality. Note that in the general case (i.e with a non-trivial NS-NS 2-form) the duality equation is modified by the presence of Chern-Simons terms. However in the case at hand, there is no NS-NS 2-form field and the duality equations are the naive ones given below. 

\subsubsection{EM duality}
As explained above, in order to perform the dualities that follow, we need to dualize the 6-form $C^{(6)}$ to a 2-form $C^{(2)}$. That is, we have to find a 
$C^{(2)}$ satisfying
\be
* dC^{(6)}=dC^{(2)} ,
\label{em}
\ee
where $*$ is performed with the metric (\ref{met}). From (\ref{C6}) we find
\be
dC^{(6)}=\Bigl[ -\frac{c_5 s_5}{H_5^2}dH\wedge (dt+c_1 c_5 \mathcal{A}) \wedge (dy+s_1 c_5 \mathcal{A}) + \frac{(1-H)s_5}{H_5}d\mathcal{A}\wedge (c_1 dy -s_1 dt) \Bigr]\wedge dz_i^4 .
\ee
Define 1-forms $\omega_1$ and $\omega_2$ as
\bea
\omega_1&=&dt +c_1 c_5 \mathcal{A},\\
\omega_2&=&dy-\frac{c_1 s_1 H}{H_1}\omega_1 +s_1 c_1 \mathcal{A} ,
\eea
so that
\be
c_1 dy-s_1 dt = c_1 \omega_{2} -\frac{s_1 (1-H)}{H_1}\omega_{1} ,
\ee
and
\be
dC^{(6)}= \left[ -\frac{c_5 s_5}{H_5^2}dH\wedge\omega_{1}\wedge\omega_{2} + \frac{(1-H)s_5}{H_5}d\mathcal{A}\wedge\left(c_1 \omega_{2} -\frac{s_1(1-H)}{H_1}\omega_{1})\right)\right]\wedge dz_i^4 .
\ee
Let $\eta^{(1)}$, $\eta^{(2)}$ be any 1 and 2-forms on $ds^2_4$. The Hodge star operation acts as
\be
* \left[ \eta^{(1)}\wedge \omega_{1} \wedge \omega_{2} \wedge dz_i^4 \right] = -\frac{H_5^2}{(1-H)^{1/2}} *_4\eta^{(1)} ,
\ee
\be
* \left[ \eta^{(2)}\wedge \omega_{1} \wedge dz_i^4 \right] = \frac{H_1 H_5}{(1-H)^{1/2}}(*_4\eta^{(2)})\wedge \omega_{2} ,
\ee
\be
* \left[ \eta^{(2)}\wedge \omega_{2} \wedge dz_i^4 \right] = \frac{H_5(1-H)^{1/2}}{H_1}(*_4\eta^{(2)})\wedge \omega_{1} .
\ee
We can use these relations to compute
\bea
* dC^{(6)}&=& c_5 s_5 \left( \frac{*_4 dH}{(1-H)^{1/2}}+(1-H)^{3/2}(*_4 d\mathcal{A})\wedge \mathcal{A}\right) \nonumber\\ && +s_5 (1-H)^{3/2}(*_4dA)\wedge (c_1 dt -s_1 dy) .
\eea
The $C^{(2)}$ solving (\ref{em}) can then be written in the form
\be
C^{(2)}=c_5 s_5 \mathcal{C}+ s_5 \mathcal{B}\wedge (c_1 dt - s_1 dy) ,
\label{c2}
\ee
where the 1-form $\mathcal{B}$ and the 2-form $\mathcal{C}$ have to satisfy
\be
d\mathcal{B} =  (1-H)^{3/2}(*_4d\mathcal{A}) ,
\label{beq}
\ee
\be
d\mathcal{C} = \left(\frac{*_4 dH}{(1-H)^{1/2}}+(1-H)^{3/2} (*_4 d\mathcal{A})\wedge \mathcal{A}\right) .
\label{ceq}
\ee
The dualization problem has thus been reduced to finding $\mathcal{B}$
and $\mathcal{C}$ that solve (\ref{beq}) and (\ref{ceq}). Note that
these equations involve only the seed vacuum metric.

Let us first look at $\mathcal{B}$. If we further decompose $\mathcal{B}$ as
\be
\mathcal{B}=\mathcal{B}_z \, (dz+\omega^1) + \mathcal{B}_\phi\,d\phi ,
\ee
we find that (\ref{beq}) implies
\be
d\mathcal{B}_z =\tau \lambda_{0a} *_3 d\omega^a
\ee 
and 
\be
d (\mathcal{B}_\phi\,d\phi) =*_3 {(\chi^{-1}d\chi)^1}_0 .
\ee
Comparing these equations with the ones defining $V_a$ and $\kappa$, we see that 
\be
\mathcal{B}_z=-V_0\,,\quad \mathcal{B}_\phi\,d\phi={\kappa^1}_0 .
\ee

Similarly let us write
\be
\mathcal{C}=(dz+\omega^1) \wedge \mathcal{C}_z ,
\ee
where $\mathcal{C}_z$ is a 1-form on the 3D base, which in our case
has only has a component along $\phi$.  
Then (\ref{ceq}) implies that
\be
d( \mathcal{C}_z+V_0 \omega^0)=*_3 {(\chi^{-1}d\chi)^0}_0 ,
\ee
so that a solution is
\be
\mathcal{C}_z=-V_0 \omega^0+{\kappa^0}_0 .
\ee

In conclusion, we have related the solution of the duality equation (\ref{em}) to the quantities $V_0$, $\omega^a$, $\kappa$ that
have been computed for the 5D vacuum solution in (\ref{v0}--\ref{k00}). The RR 2-form $C^{(2)}$ dual to $C^{(6)}$ is given by (\ref{c2}) with
\bea
\mathcal{B}&=&-V_0 (dz+\omega^1)+{\kappa^1}_0 ,\nonumber\\
\mathcal{C}&=&(dz+\omega^1) \wedge [-V_0 \omega^0+{\kappa^0}_0] .
\eea

\subsubsection{S-duality ($NS5_{y1234}-P_y$)}
\bea
ds^2_{10}&=&H_1\left[ dy -\frac{c_1 s_1 H}{H_1}(dt+c_1 c_5 \mathcal{A}) +s_1 c_5 \mathcal{A}\right]^2  \nonumber\\ &&- \frac{(1-H)}{H_1} \left[ dt+c_1 c_5 \mathcal{A} \right]^2+ H_5 ds_4^2 + ds^2_{T^4} ,\nonumber \\
B^{(2)}&=&c_5 s_5 \mathcal{C}+ s_5 \mathcal{B}\wedge (c_1 dt - s_1 dy), \\
e^{2\phi}&=&H_5 .\nonumber
\eea  
\subsubsection{T-duality along $y$ ($NS5_{y1234}-F1_y$)}

\bea
ds^2_{10}&=&H_1^{-1} \left[ dy +s_1 s_5 \mathcal{B} \right]^2 -\frac{(1-H)}{H_1} \left[ dt+c_1 c_5 \mathcal{A} \right]^2+ H_5 ds_4^2 + ds^2_{T^4} ,\nonumber \\
B^{(2)}&=&c_5 s_5 \mathcal{C}+ {c_1 s_5 \mathcal{B}\over H_1} \wedge dt +\left[ \frac{c_1 s_1 H}{H_1}(dt+c_1 c_5 \mathcal{A}) -s_1 c_5 \mathcal{A}\right]\wedge dy\\\nonumber
&&+ s_5 c_5 s_1^2 {1-H\over H_1}\mathcal{B}\wedge\mathcal{A} ,\\
e^{2\phi}&=&\frac{H_5}{H_1} .\nonumber
\eea
\subsubsection{S-duality ($D5_{y1234}-D1_y$)}
\bea
ds^2_{10}&=&H_1^{-1/2}H_5^{-1/2} \left( \left[ dy + s_1 s_5 \mathcal{B} \right]^2 -(1-H) \left[ dt+c_1 c_5 \mathcal{A} \right]^2 \right) \nonumber\\&&+  H_1^{1/2} H_5^{1/2}ds_4^2 + \frac{H_1^{1/2}}{H_5^{1/2}}ds^2_{T^4},\nonumber \\
C^{(2)}&=&c_5 s_5 \mathcal{C}+ {c_1 s_5 \mathcal{B}\over H_1} \wedge dt +\left[ \frac{c_1 s_1 H}{H_1}(dt+c_1 c_5 \mathcal{A}) -s_1 c_5 \mathcal{A}\right]\wedge dy\\\nonumber
&&+ s_5 c_5 s_1^2 {1-H\over H_1}\mathcal{B}\wedge\mathcal{A},\nonumber\\e^{2\phi}&=&\frac{H_1}{H_5}.
\label{d1d5}
\eea
This is the final result: it describes the non-extremal geometry with  $P_z$, $KK_z$, $D1_y$ and  $D5_{y1234}$ charges.

\subsection{Change of gauge}
\label{ygauge}

It is convenient for later purposes to make a coordinate
transformation
\begin{equation}
  y' =  y - s_1 s_5 \frac{Q}{P} z = y - s_1 s_5  \sqrt{ \frac{q(q^2+m^2)}{p(p^2+m^2)}}  z.
\end{equation}
If we combine this with a gauge transformation
\begin{equation}
C^{(2)} \to C^{(2)} - c_1 s_5 \sqrt{ \frac{q(q^2+m^2)}{p(p^2+m^2)}} dt
\wedge dz,
\end{equation}
this will leave the metric and two-form gauge field in the same form
as before, but with a shifted $\mathcal B$:
\begin{equation}
\mathcal B' = \mathcal B +  \sqrt{ \frac{q(q^2+m^2)}{p(p^2+m^2)}} dz =
-(V_0 - \sqrt{\frac{q(q^2+m^2)}{p(p^2+m^2)}}) (dz + \omega^1) +
(\kappa^1_0 -  \sqrt{
  \frac{q(q^2+m^2)}{p(p^2+m^2)}} \omega^1).  
\end{equation}
We would like to re-absorb this shift into a redefinition of $V_0$ and
$\kappa^1_0$ as indicated; since
\begin{equation}
\mathcal{C}=(dz+\omega^1) \wedge [-V_0 \omega^0+{\kappa^0}_0],
\end{equation}
This also involves shifting $\kappa^0_0$, 
\begin{equation}
{\kappa^0_0}' = \kappa_0^0 -  \sqrt{\frac{q(q^2+m^2)}{p(p^2+m^2)}} \omega^0.
\end{equation}
Thus, the solution is of the same form as before after this
transformation, but with the new quantities
\begin{equation} \label{V0}
V_0 = - \frac{1}{A} \sqrt{\frac{q(q^2+m^2)}{p(p^2+m^2)}} \left[ f^2 +
  2p \left(\rho + p + \frac{qb}{m} \sqrt{\frac{p^2+m^2}{q^2+m^2}} \cos
  \theta \right) \right],
\end{equation}
\begin{equation} \label{kappa10}
\kappa^1_0 = \frac{2b \sqrt{q} \sqrt{p+q}}{m \sqrt{p^2+m^2}}
\frac{\sin^2 \theta}{f^2} [\rho(pq+m^2) + m^2(p-q)], 
\end{equation}
\begin{equation} \label{kappa00}
\kappa^0_0 = -\frac{2}{f^2} q\left[ \Delta \cos \theta +
  \sqrt{\frac{q^2+m^2}{p^2+m^2}} \frac{b}{m} (p \rho-m^2) \sin^2
  \theta \right].
\end{equation}
Henceforth we will always work in this coordinate system, and will
omit the prime on $y$.

\subsection{Summary of the solution}
\label{summary}

We have now constructed an appropriate solution carrying the required
charges. Let us collect together some information about the solution
here for ease of reference. As in~\cite{nss5d}, it will be convenient for
studying the singularity structure to rewrite the solution by writing
factors of $A$ explicitly.  Let us therefore write $(1-H) = G/A$,
$H_{1,5} = \tilde H_{1,5}/A$. Then the charged metric can be written
as
\begin{eqnarray} \label{10dmet}
ds_{10}^2 &=& (\tilde H_1 \tilde H_5)^{-1/2} \left[ A(dy + s_1 s_5 
  \mathcal{B})^2 - G (dt + c_1c_5 \mathcal A)^2 \right] \\ &&+ (\tilde H_1
\tilde H_5)^{1/2} \left[ \frac{f^2}{AG} (dz + \omega^1)^2 +
  \frac{d\rho^2}{\Delta} + d\theta^2 + \frac{\Delta}{f^2} \sin^2 \theta
  d\phi^2 \right] \nonumber \\ &&+ \frac{\tilde H_1^{1/2}}{\tilde
  H_5^{1/2}} ds^2_{T^4}, \nonumber
 \end{eqnarray}
and the matter fields are
\begin{equation} \label{c2summ}
C^{(2)}=c_5 s_5 \mathcal{C}+ {c_1 s_5 A \mathcal{B}\over \tilde H_1}
\wedge dt +\left[ \frac{c_1 s_1 (A-G)}{\tilde H_1}(dt+c_1 c_5
  \mathcal{A}) -s_1 c_5 \mathcal{A}\right]\wedge dy+ s_5 c_5 s_1^2
{G\over \tilde H_1}\mathcal{B}\wedge\mathcal{A},
\end{equation}
\begin{equation}
e^{2\phi}=\frac{\tilde H_1}{\tilde H_5},
\end{equation}
where
\begin{equation} \label{af}
\mathcal A = \omega^0 - \frac{C}{G} (dz + \omega^1), 
\end{equation}
\begin{equation} \label{bf}
\mathcal B = - V_0 (dz + \omega^1) + \kappa^1_0, 
\end{equation}
\begin{equation} \label{cf}
\mathcal C = (dz + \omega^1) \wedge (-V_0 \omega^0 + \kappa^0_0) =
dz \wedge (-V_0 \omega^0 + \kappa^0_0),
\end{equation}
\begin{equation}
\tilde H_{1,5} = A + (A-G) s_{1,5}^2,
\end{equation}
\begin{equation} \label{geq}
G = A(1-H) = \frac{Af^2 - C^2}{B}.
\end{equation}
The functions from the vacuum metric are given in equations
(\ref{delta}--\ref{omega1}), and we work with the shifted $y$
coordinate, so the quantities from the electromagnetic duality are
given in (\ref{V0}--\ref{kappa00}). The determinant of the metric is
\begin{equation}
g = - \frac{\tilde H_1^3}{\tilde H_5} \sin^2 \theta. 
\end{equation}

Since both the $y$ and $z$ directions have a finite size as $\rho \to
\infty$, the solution is asymptotically flat in four dimensions. By
rearranging the metric, we can rewrite it in a form which is suitable
for Kaluza-Klein reduction,
\begin{eqnarray} \label{10dmetkk}
ds_{10}^2 &=& (\tilde H_1 \tilde H_5)^{-1/2} \left[ A(dy + s_1 s_5 
  \mathcal{B})^2 +D (dz + \omega^1 + c_1c_5 \frac{C}{D}(dt + c_1 c_5 \omega^0))^2 \right] \\ &&+ (\tilde H_1
\tilde H_5)^{1/2} \left[ -\frac{f^2}{AD} (dt + c_1 c_5 \omega^0 )^2 +
  \frac{d\rho^2}{\Delta} + d\theta^2 + \frac{\Delta}{f^2} \sin^2 \theta
  d\phi^2 \right] \nonumber \\ &&+ \frac{\tilde H_1^{1/2}}{\tilde
  H_5^{1/2}} ds^2_{T^4}, \nonumber
\end{eqnarray}
where 
\begin{equation} \label{Deq}
D = B c_1^2 c_5^2 - f^2 (c_1^2 s_5^2 + s_1^2 c_5^2) + \frac{G f^2}{A}
s_1^2 s_5^2. 
\end{equation}

The charges of the four-dimensional asymptotically flat solution are
\bea
\mathcal{M}&=& \frac{1}{2} [p+q(1+s_1 ^2 + s_5^2)],\nonumber\\
\mathcal{P}&=&P = \sqrt{\frac{ p(p^2+m^2)}{p+q}}, \nonumber\\
\mathcal{Q}&=&Q c_1 c_5 = \sqrt{\frac{q(q^2+m^2)}{(p+q)}} c_1 c_5,\nonumber \\
\mathcal{J}&=& J c_1 c_5 = b \frac{\sqrt{pq}(pq-m^2)}{m(p+q)} c_1 c_5, \nonumber\\
\mathcal{Q}_i&=&q s_i c_i, \quad i=1,5.
\label{charges}
\eea 
Here $\mathcal{M}$ is the mass of the solution and $\mathcal{J}$
its angular momentum, expressed in units for which $G_4=1$;
$\mathcal{P}$, $\mathcal{Q}$, $\mathcal{Q}_{1}$ and $\mathcal{Q}_{5}$
denote the KK monopole, KK electric, D1 and D5 charges.

\subsection{BPS limit}
\label{sectbps}

Let us consider the limit of the geometry  (\ref{10dmet}) in which $m\to 0$,  with the charges and the angular momentum held fixed. If the charges $\Qc_1$ and $\Qc_5$ are fixed to non-zero values, then the
boost parameters $\delta_1$ and $\delta_5$ must be taken to infinity. 
The resulting geometry can be parametrized 
by its charges, $\Pc$, $\Qc$, $\Qc_1$ and $\Qc_5$ (assumed to be positive), and by the
angular momentum parameter $b$, all of which are finite in this limit.  One finds, using 
Eqs. (\ref{charges}), that $p$, $q$ and $\delta_i$ should behave as
\bea
p=\Pc+O(m),\ q= \Bigl(\frac{m}{\Qc}\Bigr)^2 \frac{\Qc_1 \Qc_5}{\Pc}+O(m^3),\ \sinh\delta_i  = \frac{\Qc}{m}\sqrt\frac{\Pc \Qc_i}{\Qc_1 \Qc_5}+O(m^0),\ i=1,5.\nonumber\\
\label{extremallimit}
\eea
In this limit the mass of the solution reduces to the sum of the D1,
D5 and KK monopole charges:
\be
\mathcal{M}={1\over 2}[\Pc+\Qc_1+\Qc_5].
\ee
This shows that the limit $m\to 0$ saturates the BPS bound. 

Let us introduce the new coordinates 
\be
{\tilde r}=\rho-b\cos\theta\,,\ \cos\tilde\theta= {\rho \cos\theta-b\over \rho-b \cos\theta} .
\ee
The metric, gauge field and dilaton one obtains after performing the limit (\ref{extremallimit}) can be recast in the form
\bea
ds^2&=&(Z_1 Z_5)^{-1/2}[-(dt-k)^2+(d{y}+\omega_P - k)^2]+(Z_1 Z_5)^{1/2} ds^2_B+\Bigl({Z_1\over Z_5}\Bigr)^{1/2} ds^2_{T^4},\nonumber\\
ds^2_B&=&V^{-1} (d{z}+\chi)^2 + V (d{\tilde r}^2 + {\tilde r}^2 d{\tilde \theta}^2 + {\tilde r}^2 \sin^2{\tilde\theta}^2 d\phi^2),\nonumber\\
C^{(2)}&=&\vec{Z}_5\wedge (d{z}+\chi)+(d{y} + dt +\omega_P)\wedge \Bigl({dt+k\over Z_1}\Bigr)\nonumber\\
e^{2\Phi}&=&{Z_1\over Z_5}.
\label{extremal}
\eea
Here $Z_1$, $Z_5$, and $V$ are harmonic functions on the flat three-dimensional space spanned by the coordinates $\tilde r$, $\tilde\theta$ and $\phi$; $k$ and 
$\omega_P$ are 1-forms on the four-dimensional space with metric $ds^2_B$, of the form
\be
k= \Bigl(H_k +{H_P\over 2 V} \Bigr)(dz+\chi)+\vec{k}\,,\ \omega_P = {H_P\over V}(dz+\chi)+\vec{\omega}_P,
\ee
where $H_k$ and $H_P$ are harmonic functions and $\vec{k}$ and $\vec{\omega}$ are 1-forms on $\mathbb{R}^3$ that satisfy
\be
*_3 d\vec{k} = V d H_k -H_k dV-{d H_P\over 2}\,,\ *_3 d \vec{\omega}_P = - d H_P ,
\ee
with $*_3$ the Hodge dual on $\mathbb{R}^3$; $\chi$ and $\vec{Z}_5$ are 1-forms on $\mathbb{R}^3$ related to $V$ and $Z_5$ by
\be
*_3 d\chi = dV\,,\\ *_3d\vec{Z}_5 =  d Z_5 .
\ee
Now (\ref{extremal}) is of the general form of a supersymmetric
solution with a Gibbons-Hawking base space, and vanishing momentum
along $y$. This general form was obtained in \cite{sugrasol}. 
This shows that in the $m\to 0$ limit the solution (\ref{10dmet})
becomes supersymmetric. 

The explicit values of the functions $V$, $Z_i$, $H_k$ and $H_P$,
which are obtained by taking this limit of \eqref{10dmet} are:
\bea
&&V= 1+{Q_K\over\tilde r}\,,\ Z_i = 1+ {Q_i\over \tilde r_c}\,,\ i = 1,5\nonumber\\
&&H_k = -{Q_{Ke}\over 2 Q_K} \Bigl(1+{Q_K\over \tilde r_c}\Bigr)\, , \ H_P = {Q_{K e}\over Q_K} + {Q_1 Q_5\over Q_{K e}}\Bigl({1\over \tilde r}-{1\over\tilde r_c}\Bigr) ,
\label{extremalfunctions}
\eea
where we have defined
\bea
&&c= 2 b\,,\ Q_K = 2 \Pc\,,\ Q_{K e}= 2\Qc\,,\ Q_i = 2\Qc_i\,,\ i=1,5\nonumber\\
&&{\tilde r}_c = \sqrt{\tilde r^2 + c^2 + 2 c {\tilde r}\cos\tilde\theta} .
\eea

Let us review the analysis of the singularity structure of the
supersymmetric metric (\ref{extremal}).  A general analysis of the
regularity of metrics of the form (\ref{extremal}) has been performed
in \cite{benakraus,bw,bergl,spgp,bgl}.  One should ensure that  the
1-forms $k$ and $\omega_P$ are regular at the
point $\tilde r=0$, where the KK monopole potential $V$ diverges,. This, in particular, requires
that
\be
k_z = H_k + {H_P\over 2 V}=0
\ee 
at $\tilde r=0$. This condition is satisfied if
\be
c= {Q_K Q_{K e}^2\over Q_1 Q_5 -Q_{Ke}^2} .
\label{ccondition}
\ee
It can be checked that, with the condition (\ref{ccondition}), the
metric (\ref{extremal}, \ref{extremalfunctions}) is regular if the
coordinates $y$ and $z$ are subject to the
identifications\footnote{The metric is strictly speaking regular only
  for $N_K=1$. For $N_K$ integer greater than one, the metric has the
  usual conical singularity corresponding to $N_K$ coinciding
  monopoles.}
\bea
&&(y,z)\sim (y + 2\pi R_y, z)\sim (y ,z + 2\pi R_z)\nonumber\\
&&R_y = 2 {Q_1 Q_5\over Q_{K e}}\,,\ R_z = 2 {Q_K\over N_K}\, , \ N_K\in \mathbb{N} .
\label{extremalidentifications}
\eea
This metric with these identifications coincides with the smooth
supersymmetric D1-D5-KK solution found in \cite{benakraus} by a
completely different method, i.e. by adding KK charge to the extremal
D1-D5 geometry of \cite{bal,mm}.

\newsection{Finding smooth solutions}
\label{sectsmooth}

Within the family of metrics constructed in the previous section, we
want to see whether there are any smooth solutions. We can see by
inspection that the metric will have coordinate singularities at
$\tilde H_1= 0$, $\tilde H_5 = 0$, $\theta = 0, \pi$ and $\Delta
=0$. Because $\tilde H_{1,5}$ involve $1/B$, it will also have
singularities at $B=0$. Although the form of the metric in
\eqref{10dmet} appears to involve factors of $1/G$, these cancel out
in the actual metric coefficients, as can be seen from the alternative
form \eqref{10dmetkk}, so there is no problem at $G=0$. There is a
potential coordinate singularity at $A=0$. There is also a potential
singularity at $f^2=0$, but since $f^2 = \Delta + b^2 \sin^2 \theta$,
we will always meet a singularity at $\Delta =0$ first.

We will focus on the singularity at $\rho = \rho_0 = \sqrt{b^2 -
  m^2}$, where $\Delta =0$, and try to interpret it as a smooth
origin. As usual, $\theta = 0, \pi$ should be coordinate
singularities.  This will require appropriate identifications, to be
analysed later. We would expect that the other coordinate
singularities would be true curvature singularities, so we wish to
arrange to have solutions where $\tilde H_1, \tilde H_5, A, B >0$
everywhere. The determinant of the metric on the surfaces of constant
$\rho$ vanishes at $\Delta=0$. For the case with no momentum charge which we
are studying in this paper, we require the identifications of $y$ and
$z$ to lie in the surfaces of constant $t$. Hence for $\rho=\rho_0$ to
be a smooth origin, we need the determinant of the metric on the
surfaces of constant $\rho$ and $t$ to also vanish there. This
determinant can be easily evaluated  using \eqref{10dmetkk}:
\begin{equation}
g_{(\rho t)} = \frac{\tilde H_1^2}{\tilde H_5^2 f^2} \left[  A  \Delta D
  \sin^2 \theta - c_1^2 c_5^2 (f^2 \omega^0_\phi)^2 \right]. 
\end{equation}
In particular, at $\Delta = 0$, this is a non-zero factor times the square of
\begin{equation}
f^2 \omega^0_\phi = 2J \sin^2 \theta \left(\rho -
  \frac{m^2(p+q)}{(pq-m^2)} \right).
\end{equation}
Therefore, for the determinant to vanish at $\rho = \rho_0$, we need
\begin{equation}
\rho_0 = \sqrt{b^2 - m^2} = \frac{m^2(p+q)}{(pq-m^2)}. 
\end{equation}
This implies
\begin{equation} \label{bcond}
b^2 = \frac{m^2 (p^2+m^2)(q^2+m^2)}{(pq-m^2)^2}. 
\end{equation}
We will always assume we take the positive square root. If the
parameters satisfy \eqref{bcond}, the singularity at $\Delta =0$ is a
degeneration, where one of the spatial directions is going to zero size.

We should check that no other singularity will be encountered in the
region $\rho \geq \rho_0$, $0 \leq \theta \leq \pi$. Using
\eqref{bcond}, we can rewrite
\begin{equation}
A = f^2 + 2p \left[ (\rho - \rho_0) +
  \frac{b^2}{\rho_0} (1 + \cos \theta) \right],
\end{equation}
\begin{equation}
B = f^2 + 2q \left[ (\rho - \rho_0) +
  \frac{b^2}{\rho_0} (1 - \cos \theta) \right],
\end{equation}
so we can see that $A> 0$ and $B> 0$ for $\rho > \rho_0$. Also, 
\begin{equation}
\tilde H_{i} = Ac_{i}^2 - G s_i^2 = A \left(c_i^2 - \frac{f^2}{B}
  s_i^2 \right) + \frac{C^2}{B} s_i^2 > 0
\end{equation}
for $\rho > \rho_0$, as $A >0$ and $B > f^2$.  Thus, when
(\ref{bcond}) is satisfied, the only singularities in the metric are
at $\rho = \rho_0$ and at $\theta=0, \pi$. Each of these is a
degeneration in the $(y,z,\phi)$ part of the metric.

\subsection{Identifications}

So far, we have performed a local analysis. We now want to see what
global identifications we need to make in the $(y,z,\phi)$ space to
have a smooth metric. At each of the three coordinate singularities,
$\rho=\rho_0$, $\theta=0$, or $\theta=\pi$, some combination of these
directions is going to zero size, and we want to choose an appropriate
period to make this a smooth origin in a plane (we could in general
allow orbifold singularities, but for simplicity we focus on the task
of constructing smooth metrics). We will
write a general Killing vector in this space as $\xi = \partial_\phi -
\alpha \partial_y - \beta
\partial_z$, and choose $\alpha$ and $\beta$ to make the norm of the
Killing vector vanish at the degeneration in each case. The direction
which goes to zero size is then along $\phi$ at fixed $y + \alpha
\phi$, $z + \beta \phi$. In each case, it will turn out that we have to set
$\alpha = s_1 s_5 \kappa^1_{0,\phi}$, $\beta = \omega^1_\phi$ to make
the contributions to $\xi \cdot \xi$ from the first line in
\eqref{10dmetkk} vanish.

Consider first the singularities at $\theta = 0, \pi$. At $\theta=0$,
$f^2 = \Delta$, $\omega^0 = 0$, $\kappa^1_0 = 0$, and
\begin{equation}
\omega^1 = 2 \frac{\sqrt{p}\sqrt{p^2+m^2}}{\sqrt{p+q}} d\phi = 2
\mathcal{P} d\phi. 
\end{equation}
Thus the direction which goes to zero size at $\theta=0$ is along
$\phi$ at fixed $z + 2 \mathcal{P} \phi,$ $y$.  The metric looks
locally like $d\theta^2 + \sin^2 \theta d\phi^2$, so $\phi$ needs to
be a $2\pi$ periodic coordinate. Thus, the identification required to
make this a smooth origin is\footnote{The shift of $y$ by $z$ we
  introduced in section \ref{ygauge} was chosen to make this and
  the next identification be at constant $y$.}
\begin{equation} \label{id1} 
(y, z, \phi) \sim (y, z - 4\pi \mathcal{P}, \phi + 2\pi).
\end{equation}
Similarly, at $\theta=\pi$, $f^2 = \Delta$, $\omega^0 = 0$,
$\kappa^1_0 = 0$, and $\omega^1 = -2 \mathcal{P} d\phi$, so the
direction which goes to zero size at $\theta=\pi$ is along $\phi$ at
fixed $z - 2 \mathcal{P} \phi,$ $y$, and the required identification
is
\begin{equation} \label{id2}
(y, z, \phi) \sim (y, z + 4\pi \mathcal{P}, \phi + 2\pi). 
\end{equation}

Finally, at $\rho=\rho_0$, $f^2 = b^2 \sin^2 \theta$, $\omega^0 =0$,
$\omega^1 = -2 \mathcal{P} d\phi$,  
\begin{equation}
\kappa^1_{0,\phi} = 4 q \frac{\sqrt{q} \sqrt{p+q} }{\sqrt{q^2+m^2}},
\end{equation}
so the relevant circle is along $\phi$ at fixed $z - 2 \mathcal{P}
\phi,$ $y + 4 s_1 s_5 q \frac{\sqrt{q} \sqrt{p+q} }{\sqrt{q^2+m^2}}$.
The leading contribution to the non-zero size of this circle away from
$\rho= \rho_0$ comes just from the $\frac{\Delta}{f^2} \sin^2 \theta
d\phi^2$ term in the metric; the first line of \eqref{10dmetkk} makes
a contribution of order $(\rho -\rho_0)^2$. Therefore, writing
$\rho = \rho_0 (1 + 2 z^2)$,
the relevant part of the metric is 
\begin{equation}
4\sqrt{\tilde H_1 \tilde H_5}  ( dz^2 + \frac{\rho_0^2}{b^2}
d\phi^2). 
\end{equation}
Thus, the necessary identification here is 
\begin{equation} \label{id3}
(y, z, \phi) \sim (y -8\pi n s_1 s_5 q \frac{\sqrt{q}
  \sqrt{p+q}}{ \sqrt{q^2+m^2}} , z + 4 \pi n \mathcal{P}, \phi + 2\pi n),
\end{equation}
where $n = b/\rho_0$. We will write 
\begin{equation}
R_y = 4 q \frac{\sqrt{q}
  \sqrt{p+q}}{ \sqrt{q^2+m^2}} s_1 s_5
\end{equation}
in subsequent expressions for
compactness. We want the metric that we obtain by
Kaluza-Klein reduction from \eqref{10dmetkk} to be asymptotically flat
in four dimensions, so after the dimensional reduction, $\phi$ must be
$2\pi$ periodic. Given the identification \eqref{id3}, this imposes a
second condition on the parameters: 
\begin{equation} \label{ncond}
n = \frac{b}{\sqrt{b^2 -m^2}}  \in \mathbb{Z}.
\end{equation}
 If we consider a solution satisfying \eqref{bcond} and \eqref{ncond},
and the periodicities (\ref{id1}, \ref{id2}, \ref{id3}), the metric will
be smooth at the coordinate singularities. We will verify in the next
section that it is also smooth in the corners where two circles are
going to zero size simultaneously, and that the matter fields are
smooth. 

It is important to note that the periodicities (\ref{id1}, \ref{id2},
\ref{id3}) do not fix the lattice of identifications in the $y,z,\phi$
space uniquely. This is because although we need each of these
identifications to be a primitive vector in the lattice,\footnote{This
  is necessary to make the metric smooth at the corresponding
  coordinate singularity. If the identification is not a primitive
  lattice vector, we will have an orbifold singularity where this
  cycle degenerates.}  (\ref{id1}, \ref{id2}, \ref{id3}) do not
necessarily form a basis for the lattice. Specifying the most general
lattice consistent with the requirement that (\ref{id1}, \ref{id2},
\ref{id3}) are primitive lattice vectors is quite complicated, so we
will not discuss it in detail. As a particular example, this freedom
includes the freedom to choose the integer-quantized magnetic
Kaluza-Klein charge. We get a solution with $N_K$ units of magnetic
Kaluza-Klein charge on reduction to four dimensions by taking the basis of
identifications to be
\begin{equation} \label{Nk}
(y, x^5, \phi) \sim (y -2\pi n R_y, z, \phi) \sim (y, z +
8\pi \frac{\mathcal{P}}{N_{K}}, \phi) \sim (y, z +
4\pi \mathcal{P}, \phi +2\pi).
\end{equation}
This is one example of a large space of possibilities consistent with
(\ref{id1}, \ref{id2}, \ref{id3}). In the rest of this paper, we will
generally proceed as if (\ref{id1}, \ref{id2}, \ref{id3}) is a basis
of identifications; any other possibility corresponds to taking an
orbifold of the spacetime we describe. In particular, more general
possibilities may have orbifold singularities in the corners in the
ten-dimensional metric.

These smooth solutions admit a unique spin structure, which has antiperiodic boundary conditions for the fermions around each of the contractible cycles (\ref{id1}), (\ref{id2}) and (\ref{id3}). The fermions will thus be periodic under $z \sim z +8\pi \mathcal P$, and will be periodic under $y \sim y - 2\pi n R_y$ for odd $n$, and antiperiodic for even $n$. Thus, for odd $n$, the solutions have a spin structure compatible with preserving supersymmetry at large distances. 

\subsection{Solving the constraints}
\label{param}

There are two constraints on the parameters to obtain a smooth
solution, \eqref{bcond} and \eqref{ncond}. It is useful to have an
explicit solution of these constraints. We can obtain a simple
solution by treating $p$ and $\rho_0$ as the independent parameters,
and solving for everything else in terms of them. We then have
\begin{equation} \label{bmqpar}
b = n \rho_0, \quad m^2 = \rho_0^2 (n^2-1), \quad q = \frac{\rho_0
  (p+\rho_0) (n^2-1)}{(p - \rho_0(n^2-1))}. 
\end{equation}
We assume  $\rho_0, p$ are such that $q>0$. The various functions
appearing in the solution can be rewritten in terms of these
parameters, which makes their positivity properties more manifest:
\begin{equation}
\Delta = \rho^2 - \rho_0^2, \quad f^2 = (\rho^2-\rho_0^2) + \rho_0^2
n^2 \sin^2 \theta,
\end{equation}
\begin{equation}
A = f^2 + 2p [(\rho-\rho_0) + n^2 \rho_0 (1+\cos \theta)],
\end{equation}
\begin{equation}
B = f^2 + 2 \frac{\rho_0 (p+\rho_0) (n^2-1)}{(p - \rho_0(n^2-1))} [(\rho-\rho_0) + n^2 \rho_0 (1-\cos \theta)],
\end{equation}
\begin{equation}
C = \frac{2 \rho_0 \sqrt{ \rho_0 (\rho_0+p)} n (n^2-1)}{(p-\rho_0
  (n^2-1))} [(\rho-\rho_0) + (\rho_0+p) (1-\cos \theta)],
\end{equation}
\begin{equation}
\omega^0 = \frac{2J \sin^2 \theta (\rho-\rho_0)}{f^2} d\phi, \quad J^2
= \frac{\rho_0^3 p (\rho_0+p) n^2 (n^2-1)^2}{(p-\rho_0 (n^2-1))},
\end{equation}
\begin{equation}
\omega^1 = \frac{2}{f^2} \sqrt{p(p-\rho_0(n^2-1))} [ (\rho^2-\rho_0^2)
\cos \theta - \frac{\rho_0 p n^2}{(p - \rho_0(n^2-1))} (\rho-\rho_0)
\sin^2 \theta - n^2 \rho_0^2 \sin^2 \theta ] d\phi,
\end{equation}
and
\begin{equation}
V_0 = -\frac{n (n^2-1)}{A} \sqrt{\frac{\rho_0^3 (p+\rho_0)}{p
    (p-\rho_0(n^2-1))^3}} [ f^2 + 2p(\rho + p + (p+\rho_0) \cos
\theta)],
\end{equation}
\begin{align}
\kappa^1_0 = \frac{2 n \sqrt{\rho_0 (p+\rho_0)}}{(p-\rho_0(n^2-1))}
\frac{\sin^2 \theta}{f^2} &\left[ \frac{\rho_0(n^2-1)}{(p-\rho_0
    (n^2-1))} (p^2+2p \rho_0 - \rho_0^2(n^2-1))(\rho-\rho_0)
\right. \\ & \left. + 2
  \rho_0^2 (n^2-1) (\rho_0+p) \right], \nonumber
\end{align}
\begin{equation}
\kappa^0_0 = -\frac{2}{f^2} \frac{\rho_0 (p+\rho_0) (n^2-1)}{(p-\rho_0
  (n^2-1))} \! \left[ (\rho^2 - \rho_0^2) \cos \theta + \frac{n^2
    \rho_0}{(p-\rho_0 (n^2-1))} (p \rho - \rho_0^2(n^2-1)) \sin^2
  \theta \right].
\end{equation}
This parametrization will be used later in relating the $n=1$ case to
the supersymmetric solution and to study the near-core decoupling
limit of the solutions. 

\newsection{Verifying Regularity}
\label{reg}

\subsection{Matter fields}

The dilaton is clearly regular. For the gauge field $C^{(2)}$, we
would like to see that it is possible to make gauge transformations to
make the field regular at each of the degenerations. Recall from
section \ref{summary} that
\begin{equation}
C^{(2)}=c_5 s_5 \mathcal{C}+ {c_1 s_5 \mathcal{B}\over H_1} \wedge dt -\left[ -\frac{c_1 s_1 H}{H_1}(dt+c_1 c_5 \mathcal{A}) +s_1 c_5 \mathcal{A}\right]\wedge dy+ s_5 c_5 s_1^2 {1-H\over H_1}\mathcal{B}\wedge\mathcal{A}\\,
\end{equation}
where $\mathcal A$, $\mathcal B$, $\mathcal{C}$  are given in
(\ref{af}, \ref{bf}, \ref{cf}). 
We need to calculate the component of this two-form along the
degenerating direction $\xi = \partial_\phi - \alpha
\partial_y - \beta \partial z$. This is 
\begin{align}
\mathrm{i}_\xi C^{(2)} = &\left[ c_1 s_5 (-V_0 \omega^1_\phi +
  \kappa^1_{0,\phi}) + \alpha c_1 s_1 H + \beta c_1 s_5
  V_0 \right] \frac{dt}{H_1} \nonumber \\
&- \left[ s_1 c_5(\omega^0_\phi - \frac{C}{G}
  \omega^1_\phi) + \beta s_1 c_5\frac{C}{G} \right] \frac{(1-H)}{H_1} 
dy \nonumber \\
&+ \left[c_5 s_5 (V_0 \omega^0_\phi c_1^2 - \kappa^0_{0,\phi} H_1-(1-H)\frac{C}{G}\kappa^1_{0,\phi}s_1^2) +\alpha
  s_1 c_5 (1-H) \frac{C}{G} \right] \frac{dz}{H_1} \nonumber \\
&+\left[- \alpha s_1 c_5 (1-H) (\omega^0 - \frac{C}{G} \omega^1)
- \beta c_5 s_5 (-V_0 \omega^0 c_1^2 + \kappa^0_0 H_1 +\frac{C}{G}(1-H)\kappa^1_{0}s_1^2)\right]\frac{1}{H_1}. 
\end{align}
Since at each degeneration, $\alpha = s_1 s_5 \kappa^1_{0,\phi}$ and
$\beta = \omega^1_\phi$, we can consider the three different
degenerations simultaneously by substituting in these values of
$\alpha$ and $\beta$. Substituting these in, 
\begin{align}
\mathrm{i}_\xi C^{(2)} = & c_1 s_5 \kappa^1_{0,\phi} dt - s_1 c_5
\frac{1-H}{H_1} \omega^0_\phi dy + \left( \frac{c_5 s_5 c_1^2 V_0
    \omega^0_\phi c_1^2 }{H_1} -c_5 s_5 \kappa^0_{0,\phi} \right) dz
\\ &+
\left( \frac{s_5 c_5 (s_1^2 (H-1) \kappa^1_{0,\phi} +c_1^2 V_t
  \omega^1_\phi)\omega^0_\phi}{H_1} -c_5 s_5\omega^1_\phi
\kappa^0_{0,\phi} \right) d\phi. \nonumber
\end{align}
Note that this expression is valid only near one of the coordinate
singularities. Since $\omega^0=0$ at each of these, this expression
simplifies to
\begin{equation}
\mathrm{i}_\xi C^{(2)} = c_1 s_5 \kappa^1_{0,\phi} dt -c_5 s_5
\kappa^0_{0,\phi}  (dz + \omega^1_\phi d\phi).
\end{equation}
These remaining terms are all constants. Thus, these components of
$C^{(2)}$ are locally pure gauge, and it looks like we ought to be
able to remove them by a gauge transformation to obtain a two-form
potential which is regular at the degeneration.

However, this may not be possible globally. The integral of $C^{(2)}$
over a closed two-cycle is gauge-invariant, and if there is a non-zero
integral over a two-cycle which shrinks to zero size, it will indicate
a singularity in the gauge field. We therefore need to consider
whether there is any such integral which is non-zero. Since the
component of $C^{(2)}$ along the degenerating direction never has a
non-zero $dy$ component, the integrals to consider are where we
integrate over the degenerating cycle and one of the two cycles
\eqref{id1}, \eqref{id2}. Here we need to consider the cases
separately. If  the cycle \eqref{id1} is degenerating, then $\omega^1 = 2
\mathcal P d\phi$, and the integral over the 2-cycle formed by the product of the 1-cycles \eqref{id1} and \eqref{id2} is
\begin{equation} \label{cint}
\oint C^{(2)} = - 16\pi^2 c_5 s_5 \mathcal{P} \kappa^0_{0,\phi} =32
\pi^2 \mathcal{P} \mathcal{Q}_5. 
\end{equation}
When it is the cycle \eqref{id2} which is degenerating, $\omega^1=-2\mathcal P
d\phi$,  and the integral over the 2-cycle determined by the cycles  \eqref{id1} and \eqref{id2} has the
same value as in Eq. (\ref{cint}). When it is \eqref{id3} which is degenerating,
$\omega^1=-2\mathcal P d\phi$, so the integral over the product of \eqref{id3}
and \eqref{id2} vanishes, while the integral over \eqref{id3} and
\eqref{id1} has the same value as in (\ref{cint}).

These non-zero integrals of $C^{(2)}$ do not immediately imply a
singularity in the gauge field, as there is still the freedom to make
large gauge transformations. That is, the gauge potential (and hence
the integral) actually take values in a circle rather than the
reals. If the right-hand side of \eqref{cint} is an integer multiple
of the size of the gauge group, it can be set to zero by a large gauge
transformation.

The requirement that \eqref{cint} is an integer multiple of the size
of the gauge group is in fact just the usual quantization of a
magnetic charge, required to make the gauge field well-defined over
the whole sphere at large distance. Let us review the usual form of
this argument. The magnetic charge associated with $C^{(2)}$ is the
integral of the three-form field strength over the surface spanned by
$(\theta,z,\phi)$. If we work in a fixed gauge, we can write this
integral as
\begin{equation} 
\oint_{\theta z \phi} F^{(3)} = \oint_{z \phi} C^{(2)}|_{\theta = \pi}
- \oint_{z \phi} C^{(2)}|_{\theta = 0}. 
\end{equation}
At $\theta = 0,\pi$, the $dz \wedge d\phi$ component of $C^{(2)}$ from
\eqref{c2summ} is simply $C^{(2)}_{z\phi}|_{\theta=0,\pi} = c_s s_5 dz
\wedge \kappa^0_0 = \mp 2 q c_5 s_5 dz \wedge d\phi = \mp 2 \mathcal{Q}_5
dz \wedge d\phi$. Thus, integrating over \eqref{id1} and \eqref{id2}, 
\begin{equation} \label{magcharge}
\oint_{\theta z \phi} F^{(3)} =  32 \pi^2  \mathcal P \mathcal{Q}_5. 
\end{equation}
Now in this gauge, the gauge field is not well-behaved at either end
of the range. If we change the gauge so $C^{(2)}_{z\phi}|_{\theta=0} =
0$, then since the charge is gauge-invariant, we will have
$C^{(2)}_{z\phi}|_{\theta=\pi} = 4 \mathcal{Q}_5 dz \wedge d\phi$. For
the gauge field to be globally well-behaved on the whole surface,
there must be a large gauge transformation which can be used to shift
this to zero. This is equivalent to requiring that the charge
\eqref{magcharge} is a multiple of the size of the gauge group. This
large gauge transformation is then precisely what we need to see that
the integral \eqref{cint} of the two-form over the degenerating
two-cycles is gauge-equivalent to zero. Thus, we
have succeeded in showing that the gauge field is regular up to gauge
transformations.

\subsection{Corners}

With the conditions above, the solution is smooth at $\rho = \rho_0$
or $\theta = 0, \pi$. However, it is not clear what happens in the
`corners', where $\rho = \rho_0$ and $\theta = 0, \pi$. In this
section, we will introduce coordinates which explicitly show that the
ten-dimensional geometry is smooth at these points as well. 

Consider first the corner at $\rho = \rho_0$, $\theta =0$. Define new
coordinates by\footnote{Note that for $n=1$, these are the same as the
  coordinates used in section \ref{sectbps}.}
\begin{equation}
\tilde r = (\rho-\rho_0) + \rho_0 (1-\cos
\theta),
\end{equation}
\begin{equation}
\tilde r \cos^2 \frac{\tilde \theta}{2} = 
\frac{\rho-\rho_0}{2} (1 + \cos
\theta).
\end{equation}
In the new coordinates, $\rho = \rho_0$, $\theta =0$ is at $\tilde
r=0$, with $\rho = \rho_0$, $\theta \neq 0$ along $\tilde \theta
=\pi$, and $\rho \neq \rho_0$, $\theta =0$ along $\tilde \theta
=0$. In these coordinates,
\begin{equation}
\frac{d\rho^2}{\Delta} + d\theta^2 = \frac{1}{\tilde r \tilde r_c}
(d\tilde r^2 + \tilde r^2 d\tilde \theta^2), 
\end{equation}
with $\tilde r_c^2 = \tilde r^2 + 4 \tilde r \rho_0 \cos \tilde \theta +
4 \rho_0^2$. Near $\tilde r=0$, 
\begin{equation}
\rho -\rho_0 \approx \frac{\tilde r}{2} (1 + \cos \tilde
  \theta), \quad \sin^2 \theta \approx \frac{ \tilde r}{\rho_0} (1
- \cos \tilde \theta),
\end{equation}
so
\begin{equation}
\Delta \approx \tilde r \rho_0 (1 + \cos \tilde \theta),
\end{equation}
\begin{equation}
f^2 \approx  2\tilde r \rho_0 \gamma ,
\end{equation}
where  
\begin{equation}
  \gamma = \frac{1}{2}[(1+ \cos \tilde \theta) +
  n^2 (1 - \cos \tilde \theta) ].
\end{equation}
We also have
\begin{equation}
A \approx \frac{4p
  b^2}{\rho_0},
\end{equation}
\begin{equation}
B \approx 2\tilde r (\rho_0+q) \gamma,
\end{equation}
\begin{equation}
C \approx \tilde r \frac{\sqrt{q} \sqrt{q^2+m^2}}{\sqrt{p+q}}
\left[ (1 + \cos \tilde \theta) + \frac{\rho_0+p}{\rho_0} (1 - \cos
  \tilde \theta) \right].
\end{equation}
The above scalings imply that $G$, and hence $\tilde H_{1,5}$, remain
finite as $\tilde r \to 0$: the vanishing of $B$ in the denominator of
\eqref{geq} is cancelled by the factor of $A$, $C^2$ in the numerator.
Also, $\tilde H_{1,5}$ are constants, as
\begin{equation}
G \approx \frac{Af^2}{B} \approx \frac{A \rho_0}{\rho_0 +q}.  
\end{equation}
The one-form ${\mathcal A} \sim {\mathcal O}(\tilde r)$, so we can
ignore it, while
\be
\mathcal{B}\approx 2 q \sqrt{q (p+q)\over q^2+m^2} \left( d\phi +
  \frac{dz}{2\mathcal P} \right).
\ee
Hence the first line in \eqref{10dmet} just involves constants in this
limit. After some algebra, the non-constant part of the metric becomes
\begin{align}
 &\frac{f^2}{AG} (dz + \omega^1)^2+ \frac{d\rho^2}{\Delta} +
  d\theta^2 + \frac{\Delta}{f^2} \sin^2 \theta d\phi^2 \approx
  \frac{1}{2 \rho_0 \tilde r} \left[ d \tilde r^2 + \tilde r^2 d\tilde
    \theta^2 \right. \\ &+ \left. 
  \frac{\tilde r^2}{2 n^2} (1+\cos \tilde \theta) \left( d\phi +
    \frac{dz}{2\mathcal P} \right)^2+
  \frac{\tilde r^2}{2} (1-\cos \tilde \theta) \left( d\phi -
    \frac{dz}{2 \mathcal P} \right)^2 \right]. \nonumber 
\end{align}
Thus, if we define coordinates $\tilde r = R^2$, $\tilde \theta_c = 2
\vartheta$,
\begin{equation}
\psi_1 = \frac{1}{2}\left(\phi - \frac{z}{2 \mathcal P} \right), \quad \psi_2 = \frac{1}{2n}
\left( \phi + \frac{z}{2 \mathcal P} \right), 
\end{equation}
\begin{equation}
\hat y = y +\frac{R_y}{2} \left( \phi +
  \frac{z}{2 \mathcal P} \right), 
\end{equation}
the non-constant part of the metric becomes the standard metric on
$\mathbb R^4$, while the identifications
(\ref{id1}, \ref{id2}, \ref{id3}) become respectively $\psi_1 \sim
\psi_1 +2 \pi$, $\hat y \sim \hat y + 2\pi R_y$, $\psi_2 \sim \psi_2
+2\pi$. Thus, the local geometry is globally $\mathbb R^4$, and hence
smooth near this corner.

Consider next the corner at $\rho=\rho_0$, $\theta = \pi$. We
similarly define new coordinates
\begin{equation}
\tilde r_c =  (\rho-\rho_0) + \rho_0 (1+\cos
\theta),
\end{equation}
\begin{equation}
\tilde r_c \cos^2 \frac{\tilde \theta_c}{2} = 
\frac{\rho-\rho_0}{2} (1 - \cos
\theta).
\end{equation}
In the new coordinates, $\rho = \rho_0$, $\theta =\pi$ is at $\tilde
r_c=0$, with $\rho = \rho_0$, $\theta \neq \pi$ along $\tilde \theta_c
=\pi$, and $\rho \neq \rho_0$, $\theta =\pi$ along $\tilde \theta_c
=0$.  In these coordinates,
\begin{equation}
\frac{d\rho^2}{\Delta} + d\theta^2 = \frac{1}{\tilde r \tilde r_c}
(d\tilde r_c^2 + \tilde r_c^2 d\tilde \theta_c^2), 
\end{equation}
where now $\tilde r^2 = \tilde r_c^2 + 4 \rho_0 \tilde r_c  \cos
\tilde \theta_c + 4 \rho_0^2$. Near $\tilde r_c =0$, 
\begin{equation}
\rho -\rho_0 \approx \frac{\tilde r_c}{2} (1 + \cos \tilde
  \theta_c), \quad \sin^2 \theta \approx \frac{ \tilde r_c}{\rho_0} (1
- \cos \tilde \theta_c). 
\end{equation}
so 
\begin{equation}
\Delta \approx \tilde r_c \rho_0 (1 + \cos \tilde \theta_c),
\end{equation}
\begin{equation}
f^2 \approx  2 \tilde r_c \rho_0 \gamma_c ,
\end{equation}
where as before we will define 
\begin{equation}
  \gamma_c = \frac{1}{2}[(1+ \cos \tilde \theta_c) +
  n^2 (1 - \cos \tilde \theta_c) ].
\end{equation}
We also have
\begin{equation}
A \approx 2\tilde r_c (\rho_0+p) \gamma_c,
\end{equation}
\begin{equation}
B \approx \frac{4q b^2}{\rho_0},
\end{equation}
\begin{equation}
C \approx \frac{4 q^{3/2} \sqrt{q^2+m^2} (p^2+m^2)}{\sqrt{p+q} (pq-m^2)}.
\end{equation}
Thus for small $\tilde r_c$, $G \approx -C^2/B$ is a constant, and
$\tilde H_{1,5}$ are then constants:
\begin{equation}
\tilde H_{1,5} \approx \frac{C^2}{B} s^2_{1,5} \approx \frac{4 q^2
  (p^2+m^2)}{(pq-m^2)} s^2_{1,5}. 
\end{equation}
Also, 
\begin{equation}
\frac{f^2}{AG} \approx - \frac{\rho_0}{\rho_0+p}, \quad \omega^1
\approx -2\mathcal P d\phi,
\end{equation}
so the $\frac{f^2}{AG} (dz+ \omega^1)^2$ term in the metric is a constant
size circle, and the non-constant part of the metric is, up to an
overall factor, 
\begin{equation}
d\Sigma^2 = \frac{d\rho^2}{\Delta} + d\theta^2 + \frac{\Delta}{f^2}
\sin^2 \theta d\phi^2 + \frac{A}{\tilde H_1 \tilde H_5} (dy - s_1 s_5
V_0 (dz + \omega^1) + s_1 s_5 \kappa^1_0)^2. 
\end{equation}
After considerable algebra, this becomes
\begin{eqnarray}
d\Sigma^2 &\approx& \frac{1}{2\rho_0 \tilde r_c} (d \tilde r_c^2 + \tilde r_c^2 d
\tilde \theta_c^2) \\
&&+ \frac{\tilde r_c}{ \rho_0} (1+ \cos \tilde
  \theta_c) \frac{1}{n^2 R_y^2} \left( d y - s_1 s_5
V_0 (dz + \omega^1) 
\right)^2 \nonumber \\
&&+ \frac{\tilde r_c}{\rho_0} (1- \cos \tilde
  \theta_c)  \left(
   d\phi + \frac{1}{R_y} (dy - s_1 s_5
V_0 (dz + \omega^1)) \right)^2. \nonumber 
\end{eqnarray}
If we set $\tilde r_c = R^2$, $\tilde \theta_c = 2
\vartheta$, $\tilde z = z - 2\mathcal P \phi$, 
\begin{equation}
\psi_1 = \left( \phi+ \frac{1}{R_y} y \right), \quad \psi_2 =
-\frac{1}{n R_y} y,
\end{equation}
this becomes the standard metric on $\mathbb R^4$, plus some terms
involving $\tilde z$ which are small compared to the $\frac{f^2}{AG}d
\tilde z^2$ factor.  The identifications \eqref{id2}, \eqref{id3}
become in these coordinates simply $\psi_1 \sim \psi_1 + 2\pi$,
$\psi_2 \sim \psi_2 + 2\pi$, so the metric is globally $\mathbb R^4$,
and hence smooth near this corner.

\subsection{Closed timelike curves}

Finally, we  verify the absence of closed timelike curves in this metric. We will
do this by showing that $t$ is a global time function, which requires
$g^{tt} < 0$ everywhere.  A basis of orthonormal vector fields for
\eqref{10dmetkk} is
\begin{equation}
e^1 = \frac{(\tilde H_1 \tilde H_5)^{1/4}}{\sqrt{A}} \partial_y, \quad
e^2 = \frac{(\tilde H_1 \tilde H_5)^{1/4}}{\sqrt{D}} (\partial_z + s_1
s_5 V_0 \partial_y),
\end{equation}
\begin{equation}
e^3 = \frac{\sqrt{\Delta}}{(\tilde H_1 \tilde H_5)^{1/4}} \partial_r, \quad
e^4 = \frac{1}{(\tilde H_1 \tilde H_5)^{1/4}} \partial_\theta,
\end{equation}
\begin{equation}
e^0 = \frac{\sqrt{AD}}{f(\tilde H_1 \tilde H_5)^{1/4}}( \partial_{t}
- c_1 c_5 \partial_z ),
\end{equation}
\begin{equation}
e^5 = \frac{f}{\sqrt{\Delta} \sin \theta (\tilde H_1 \tilde H_5)^{1/4}}
( \partial_\phi - s_1 s_5
\kappa^1_{0,\phi} \partial_y - \omega^1_\phi \partial_{z} - c_1 c_5 \omega^0_\phi \partial_t), 
\end{equation}
plus four more for the $T^4$. From this, we can compute
\begin{equation} \label{gtt}
  g^{tt} = - \frac{AD}{f^2\sqrt{\tilde H_1 \tilde H_5}} + \frac{f^2}{\Delta \sin^2
    \theta \sqrt{\tilde H_1
      \tilde H_5}} c_1^2 c_5^2 (\omega^0_\phi)^2.
\end{equation}

To show this is negative, we will write it in terms of separate factors
independent of the charges, and show that each of the factors
 is negative separately. Let us write $g^{tt} = - \frac{1}{f^2 \Delta
  \sqrt{\tilde H_1 \tilde H_5}} U$, where
\be
U=F_1 + (s_1^2+s_5^2)F_2 +c_1^2 c_5^2 F_3
\ee
and
\bea
&&F_1 = (1+H) A f^2 \Delta, \quad F_2 = H A f^2 \Delta\label{2terms},\\
&&F_3 =A(B-f^2)\Delta-H A f^2 \Delta -  4J^2 \sin^2 \theta
(\rho-\rho_0)^2. \label{messyterm}
\eea
We already know that $A$, $B$, $f^2$ and $\Delta$ are positive for $\rho>\rho_0$. It is also easy to see
that $H$ is positive, as
\be
H ={(B-f^2)\over B}+{C^2\over A B}>0
\ee
because $B>f^2$. This implies that $F_1, F_2 >0$. It remains to be shown 
that $F_3 >0$.

To show that this last term is also positive, we rewrite it as 
\be
F_3 = \frac{S}{B} = \frac{1}{B} \left[ ( A(B-f^2)^2 - C^2  f^2 ) \Delta - B J^2 (\rho-\rho_0)^2 \sin^2\theta \right].
\ee
As we already know that $B>0$, it is sufficient to show that the term
$S$ is positive. This term is a sixth order polynomial in $r=
(\rho-\rho_0)$,
\be
 S = c_6 r^6 + c_5 r^5 + c_4 r^4 + c_3 r^3 + c_2 r^2 + c_1 r + c_0.
\ee
To prove that $S\geq0$ it is sufficient (though not necessary) to show that the individual coefficients $c_i$ are positive. We find
\bea
c_6 &\!\!=\!\!&  \frac{4 q(p q-m^2)}{p+q},\\
c_5 &\!\!=\!\!& 8 q (p q + 2 m^2),  \\
c_4 &\!\!=\!\!&c_{40} + c_{41}\cos\theta + c_{42} \cos^2\theta,  \\
c_3&\!\!=\!\!& c_{30} + c_{31}\cos\theta + c_{32} \cos^2\theta,  \\
c_2 &\!\!=\!\!& \frac{4 m^2 (1-\cos\theta)q(p^2+m^2)^2(q^2+m^2)}{(p+q)^2 (p q-m^2)^3} \left[ c_{20} + c_{21}\cos\theta + c_{22} \cos^2\theta +c_{23} \cos^3\theta \right],  \nonumber \\ \label{coeff2} \\
c_1 &=& \frac{8 m^2 q^2\sin^2\theta (1-\cos\theta) (p^2+m^2)^3 (q^2+m^2)^2 [ m^2(p+q) (1+\cos\theta  ) +2 q (p q-m^2) ] }{(p+q)^2 (p q -m^2)^4 },\nonumber \\ \label{c1}
 \\
c_0 &\!\!=\!\!& 0.
\eea
In the following we will not need the explicit values of the coefficients $c_{2i}$, $c_{3i}$ and $c_{4i}$.
Noting that $p>0,q>0$ and $p q-m^2>0$, the first two coefficients are
immediately seen to be positive. For $c_4$, $c_3$ and $c_2$ the story
is more complicated, and it turned out to be simplest to show these
terms are positive indirectly. Let $x\equiv \cos\theta$. Then for
$c_4$ we have
\bea
c_4(-1)& =& (p+q)^{-2}  (pq -m^2)^{-1} \left[16 p q ((p^2+3 q p+4 q^2)
  m^4\right.  \\ & & \nonumber \left. +q (2 p^3+5 q p^2+3 q^2 p+3 q^3) m^2 +p^2 q^3 (p+2 q))+16 m^4 q^2 \left(p q-m^2\right)\right]>0, \\
c_4(1) &=& \frac{16 q}{(p+q)(pq -m^2)} \left[\left(p^2+3 q
    p+q^2\right) m^4+2 p q \left(p^2+q p+q^2\right) m^2
\right. \nonumber \\ && \left. +p^3 q^3\right]>0, \\
c_4''(x) &=& -\frac{16 m^2 q (p^2+m^2)(q^2+m^2)}{(p+q) (p q -m^2)}<0.
\eea
From this data one can see that $c_4$ is positive at the boundaries,
and is an inverted parabola. This implies that $c_4(x) > 0$ for all $x$
between $-1$ and $1$. Similar data for $c_3$ also proves that it is
positive for all values of $\theta$:
\bea
c_3(-1)& =& \frac{32q^2 (p^2+m^2)(q^2+m^2)}{(p+q)^{2}  (pq -m^2)^{2}}
\left[p^3 q^2+ m^2 p(p^2+ p q + q^2) \right. \nonumber \\ && \left.+ m^2(p+2q)(p q -m^2)  \right] >0,
\\
c_3(1) &=& \frac{32 m^2 p q^2 (p^2+m^2)(q^2+m^2) }{  (pq -m^2)^{2}}>0,  \\
c_3''(x) &=& -\frac{16 m^2 q(p^2+m^2)(q^2+m^2)(p q(2q+p)+m^2(2p+3q))}{(p+q) (p q -m^2)^2} <0.
\eea
For $c_2$ the argument is more complicated. The prefactors in
(\ref{coeff2}) are clearly positive so it is only necessary to
consider the bracketed term. Let us call this term $\tilde{c}_2$.  We
first prove that $\tilde{c}_2'(x)$ is positive for $x\in [-1,1]$. This
we do by furnishing the same data as done above for $c_4$ and $c_3$:
\bea
\tilde{c}_{2}'(-1) &=& 4 q (pq^2 (p+2 q) + p m^2 (2 p + 3 q) + m^2(p
q-m^2)) >0, \\
\tilde{c}_{2}'(1) &=&4 q \left[\left(p^2+p q +q^2\right) m^2+p^2
  q^2\right] + 4 m^2 p (p q-m^2) >0, \\
\tilde{c}_{2}''' &=& - 6 m^2 (p+q)(q^2+m^2)< 0.
\eea
Now given that $\tilde{c}_2'(x)$ is positive for $x\in [-1,1]$ it is
sufficient to show that $\tilde{c}_2(-1)>0$ in order to prove that
$\tilde{c}_{2}$ is positive for all $\theta$. We find
\be
\tilde{c}_{2}(-1) =\frac{64 m^2q^3 (p^2+m^2)^2(q^2+m^2)}{(p +q)(p q-m^2)^2} >0.
\ee
Finally for $c_1$ it is clear from (\ref{c1}) that it is positive
for all $\theta$. Thus we have shown that $S$ is positive for all
values of $\rho >\rho_0$ and $\theta$. Hence $g^{tt} <0$, so $t$ is a
global time function, and there are no closed timelike curves in the geometry.

Note also from \eqref{gtt} that $g^{tt} <0$ implies $D>0$. This will
be significant in the analysis of the four-dimensional solution. 

\newsection{Properties of the solutions}
\label{sectprops}

\subsection{The supersymmetric case}

In the special case $n=1$, we would expect to recover the
supersymmetric solution of \cite{benakraus}. From the parametrization
\eqref{bmqpar}, we can see that $m \to 0$ with $q/m^2$ and $b$ fixed as $n
\to 1$. Thus, if we scale $\delta_i \to \infty$ so as to keep $m^2 s_i
c_i$ fixed as we take $n \to 1$, we will be taking the extremal limit
described in (\ref{extremallimit}).  In this limit the constraint
\eqref{ncond} reduces to the regularity condition \eqref{ccondition}
we have found in Section \ref{sectbps} for the supersymmetric
solution. Also the identifications (\ref{Nk}) become equivalent to
(\ref{extremalidentifications}). Thus one can think of the regular
supersymmetric geometry (\ref{extremal}, \ref{extremalfunctions}) as
the particular member of the class of smooth metrics of Section
\ref{sectsmooth} with $n=1$, provided that one also takes the
$\delta_i$ parameters to infinity, as specified in
(\ref{extremallimit}).

\subsection{Ergoregion}

In the five-dimensional solutions studied in \cite{nss5d}, one of the
most striking and important properties of the solutions was that they
have an ergoregion, where the timelike Killing vector at infinity
becomes spacelike. The existence of an ergoregion is a characteristic
property of the non-supersymmetric solutions: unbroken supersymmetry,
by contrast, implies the existence of an everywhere causal Killing
vector. In~\cite{Cardoso}, this ergoregion was also shown to imply
that the non-supersymmetric solutions of~\cite{nss5d} were unstable,
using a general argument due to~\cite{Friedmann}. It is therefore
clearly important to study the ergoregion in our solutions.

It is difficult to analyse the ergoregion in the ten-dimensional
geometry \eqref{10dmet}. The most general Killing vector which is
timelike at large $\rho$ is a linear combination of $\partial_t$,
$\partial_y$, and $\partial_z$, $\xi = \partial_t - a \partial_y -
b \partial_z$. We have
\begin{equation}
\xi \cdot \xi = \frac{1}{\sqrt{\tilde H_1 \tilde H_5}} \left[ A (a -
  s_1 s_5 V_0 b)^2 -G + 2 C c_1 c_5 b + D b^2 \right]. 
\end{equation}
Requiring this to be timelike at large $\rho$ imposes 
\begin{equation} \label{easymp}
\left( a + s_1 s_5  \sqrt{ \frac{q(q^2+m^2)}{p(p^2+m^2)}} b \right)^2
+ b^2 < 1.
\end{equation}
The expression for $\xi \cdot \xi$ is complicated. To get some
insight, we can examine its behaviour in the corners: at $\rho =
\rho_0, \theta=0$,
\begin{equation} \label{ecorner1} 
\xi \cdot \xi = \frac{1}{\sqrt{\tilde H_1 \tilde H_5}} \left[ 4p
  \rho_0 n^2 (a - V_0 s_1 s_5 b)^2 - 4 \rho_0 (p - \rho_0 (n^2-1)) \right], 
\end{equation}
where
\begin{equation}
V_0(\rho=\rho_0, \theta =0) = - (n^2-1) \sqrt{ \frac{\rho_0
    (p+\rho_0)^3}{n^2 p (p-\rho_0(n^2-1))}}. 
\end{equation}
At $\rho = \rho_0, \theta=\pi$, 
\begin{equation} \label{ecorner2}
\xi \cdot \xi = \frac{4 \rho_0 (\rho_0 +p)^2 (n^2-1)}{(p-\rho_0
  (n^2-1))\sqrt{\tilde H_1 \tilde H_5}} [1 + (n c_1 c_5 - s_1 s_5)
b][1 + (n c_1 c_5 + s_1 s_5) b]. 
\end{equation}
A necessary condition for $\xi$ to be everywhere timelike in the
ten-dimensional geometry is that we can choose $a$ and $b$ to make
\eqref{ecorner1} and \eqref{ecorner2} negative while satisfying
\eqref{easymp}. We have not analysed these conditions in detail; they
depend in a complicated way on the charges. 

Instead, we will study the ergoregion in the four-dimensional metric
we obtain by Kaluza-Klein reduction. The ergoregion in the
four-dimensional metric is in general different from the ergoregion in
the ten-dimensional metric, since in the Kaluza-Klein reduction, we
project the Killing vector down to four dimensions, losing the
contribution to its norm from the first line in \eqref{10dmetkk}. The
instability of~\cite{Friedmann,Cardoso} was determined by the presence
of an ergoregion in the asymptotically flat metric, so the ergoregion
in the four-dimensional metric would seem to be more relevant to the
question of stability. It also turns out to be much easier to
determine. For this, we use \eqref{10dmetkk}, where the
four-dimensional metric is given, up to a conformal factor, by the
second line. But we have shown that $f>0$, $A>0$, and $D>0$ away from the
degenerations, so in this 4d metric, $\partial_t$ is timelike
everywhere. Thus, there is no ergoregion in the 4d metric!

This might seem quite surprising, but we can understand the difference
from the five-dimensional case on general grounds, without detailed
calculation. The four-dimensional metric we obtain upon Kaluza-Klein
reduction is given by \eqref{10dmetkk} for some $D$. Now for this to
have an ergoregion, we would need $g_{tt}$ to change sign while the
four-dimensional metric remains of fixed signature. If we think of the
second line of \eqref{10dmetkk} as the $t$ direction fibred over a
three-dimensional base metric, to preserve the overall signature, the
determinant of the base metric would have to change sign. But these
terms are clearly all everywhere positive: in particular, the factor
in front of $d\phi^2$ is positive away from the degenerations. The
difference in the five-dimensional case was that we had a pair of
angular directions, so the determinant of the four-dimensional base
metric could change sign without encountering any degenerations. Thus,
we expect the absence of the ergoregion in the four-dimensional
solution to be a general property of such solutions.

Thus, the Killing vector $V = \partial_t$ is timelike everywhere in
the four-dimensional spacetime. Assuming that we consider test fields
propagating on this spacetime which satisfy the dominant energy
condition, it follows that the energy constructed by integration over
a Cauchy surface,
\begin{equation}
\mathcal E = \int_S V^\mu T_{\mu}^{\ \nu} dS_\nu,
\end{equation}
will always be positive for any initial data. Hence, the instability
discussed in~\cite{Friedmann,Cardoso} cannot arise in this case. It is
then an open question whether our non-supersymmetric solutions are
unstable. There is no mechanism that would prevent them from being
unstable, so past experience biases us to think that they will be, but
this is a very interesting question for future research.

\subsection{Near-core limit}

The solutions we have constructed look qualitatively like smooth D1-D5
solutions sitting at the core of a Kaluza-Klein monopole. We would
therefore expect to find that there is a suitable decoupling limit of
the geometry in which we focus on the core region, and obtain an
AdS$_3 \times S^3$ geometry. As in the previous non-supersymmetric
case~\cite{nss5d}, obtaining such a limit will require us to scale
some of the charges in a suitable way, going close to extremality. In
this section, we will construct the decoupling limit for these
solutions.

In the parametrization of section~\ref{param}, the only free parameters
are $p, \rho_0$ and the charge parameters. It seems natural to
consider a limit where we take $\rho_0 \to 0$, while holding $p$ and the
physical D1 and D5 charges fixed: that is, we hold $\rho_0 \sinh
\delta_i \cosh \delta_i$ fixed. Note that this is {\emph not} the same
as the extremal limit introduced in section \ref{sectbps}, in
which we took $m \to 0$ holding $b$ fixed. In fact, such a limit is
incompatible with the constraints imposed by the smoothness
conditions. Thus, here we are not taking the extremal limit with all
the charges held fixed; instead, we are scaling $\mathcal Q$ and
$\mathcal J$ to zero.

As we take this limit, we scale the coordinates so as to zoom in on a
`core' region in the geometry, by setting $\rho = \rho_0 r$ and
holding $r$ fixed. As we take the limit, the identification on the $y$
coordinate scales like $1/\sqrt{\rho_0}$. It is therefore convenient
to set $y = \chi/ 4\sqrt{p \rho_0}$. It will also be convenient to set
$t = \tau/ 4\sqrt{p \rho_0}$ and $z=p \psi$. In this limit, the metric
\eqref{10dmet} becomes
\begin{align} \label{ads1}
ds_{10}^2 \approx& \frac{1}{4\ell^2} \left\{ a \left[ d\chi +
    \frac{\ell^2 n}{2} \left( \frac{(1+\cos \theta)}{a} (d\psi + \bar
      \omega^1) + \bar \kappa^1 \right) \right]^2 \right. \\ 
& \left. -g \left[ d \tau + \frac{\ell^2 n}{2} \left( \bar \omega^0 -
    \frac{(1-\cos \theta)}{g} (d\psi + \bar \omega^1) \right)
\right]^2 \right\} \nonumber \\
& + \frac{\ell^2}{4} \left[ \frac{dr^2}{r^2-1} + d\theta^2 + \frac{r^2-1+n^2
    \sin^2 \theta}{ag} (d\psi + \bar \omega^1)^2 + \frac{(r^2-1)
    \sin^2 \theta}{(r^2-1+n^2 \sin^2 \theta)} d\phi^2 \right]
\nonumber \\ 
&+ \sqrt{\frac{\mathcal Q_1}{\mathcal Q_5}} ds^2_{T^4}, \nonumber
\end{align}
where we have set 
\begin{equation}
\ell^2 = 4\sqrt{\tilde H_1 \tilde H_5} = 16p
\sqrt{\mathcal Q_1 \mathcal Q_5},
\end{equation}
and
\begin{equation}
a = 2(r-1+n^2(1+\cos \theta)), \quad g = 2(r+1-n^2 (1-\cos \theta)), 
\end{equation}
\begin{equation}
\bar \omega^0 = \frac{(r-1) \sin^2 \theta}{r^2-1+n^2 \sin^2 \theta}
d\phi,
\end{equation}
\begin{equation}
\bar \omega^1 = 2\frac{(r^2-1) \cos \theta - n^2 r^2 \sin^2
  \theta}{r^2-1+n^2 \sin^2 \theta} d\phi,
\end{equation}
\begin{equation}
\bar \kappa^1 = \frac{(r+1) \sin^2 \theta}{r^2-1+n^2 \sin^2 \theta}. 
\end{equation}
This metric has an AdS$_3 \times S^3$ geometry (at least
locally). This can be made explicit by introducing new angular
coordinates 
\begin{equation}
\bar \psi = \frac{1}{4} (2\phi + \psi), \quad \bar \phi =
\frac{1}{4}(2\phi - \psi),
\end{equation}
and writing
\begin{equation}
r = 1 + 2 R^2, \quad \chi = \ell^2 \varphi, \quad \theta = 2\bar \theta.
\end{equation}
In terms of these coordinates \eqref{ads1} becomes
\begin{align}
ds^2 = &-\frac{(R^2+1)}{\ell^2} d\tau^2 + \frac{\ell^2 dR^2}{R^2+1}
+ \ell^2 R^2 d\varphi^2 \\ &+ \ell^2 (d\bar \theta^2 + \cos^2 \bar
\theta (d\bar \psi + n d\varphi)^2 + \sin^2 \bar \theta (d\bar \phi -
\frac{n}{\ell^2} d\tau)^2)+ \sqrt{\frac{\mathcal Q_1}{\mathcal Q_5}} ds^2_{T^4}. \nonumber
\end{align}
The identifications (\ref{id1}, \ref{id2}, \ref{id3}) become in these
coordinates simply $\bar \psi \sim \bar \psi + 2\pi$, $\bar \phi \sim
\bar \phi + 2\pi$ and $(\varphi, \bar \psi) \sim (\varphi - 2\pi, \bar
\psi + 2\pi n)$. These identifications make the spacetime globally
AdS$_3 \times S^3$. Recall however that these may not be the
fundamental identifications. For example, if we adopt the basis of
identifications \eqref{Nk}, the geometry becomes AdS$_3 \times
S^3/\mathbb{Z}_{N_K}$. More general choices will give other orbifolds
of AdS$_3 \times S^3$ in the decoupling limit.

The dilaton is a constant in this limit. For the form field, we first
need to make a gauge transformation to get the correct behaviour in
the limit: we shift $C^{(2)} \to C^{(2)} - dt \wedge dy$, so that
\begin{eqnarray}
C^{(2)} &= & c_5 s_5 \mathcal{C}+ s_5 c_5 s_1^2 {1-H\over
  H_1}\mathcal{B}\wedge\mathcal{A} + {c_1 s_5 \mathcal{B}\over H_1}
\wedge dt - \frac{s_1 c_5 \mathcal{A}(1-H)}{H_1} \wedge dy \\ && - \frac{(1 -
s_1(c_1-s_1) H)}{H_1} dt \wedge dy. \nonumber
\end{eqnarray}
Now as $\rho_0 \to 0$, this will become 
\begin{align} \label{climit}
C^{(2)} = & p\mathcal Q_5 \left( \frac{\mathcal{C}}{p \rho_0 (n^2-1)} + \frac{16 n^2
    \sin^2 \theta}{a} d\bar \psi \wedge d\bar \phi \right) + \sqrt{
  \frac{\mathcal Q_5}{\mathcal Q_1}} \frac{n}{2} (1+\cos \theta) d\bar \psi \wedge d\tau \\ &-\sqrt{
  \frac{\mathcal Q_5}{\mathcal Q_1}} \frac{n}{2} (1-\cos \theta) d\bar \phi \wedge d\chi -
  \frac{1}{32p \mathcal Q_1} (r+1-n^2 (1-\cos \theta)) d\tau \wedge d \chi. \nonumber
\end{align}
In the limit,
\begin{equation}
\mathcal C = dx^5 \wedge ( -V_t \omega^0 + \kappa^0_t) \approx -16
\rho_0 p(n^2-1) \frac{((r-1) \cos \theta + n^2(1+\cos \theta))}{a}
d\bar \psi \wedge d \bar \phi,
\end{equation}
so discarding some pure gauge terms from \eqref{climit}, the two-form becomes
\begin{equation}
C^{(2)} = -8p \mathcal{Q}_5 \left[ \cos \theta (d\bar
    \psi + \frac{n}{\ell^2} d\chi) \wedge (d \bar \phi -
    \frac{n}{\ell^2} d\tau) + \frac{1}{\ell^4} r d\tau \wedge d \chi \right],
\end{equation}
which is of the expected form to correspond to an AdS$_3 \times S^3$ solution.

\subsection{Four-dimensional description}

Finally, we will make a brief remark about the structure of the
four-dimensional metric obtained by Kaluza-Klein reduction. The
four-dimensional metric in the Einstein frame is
\begin{equation}
ds^2_4 = -\frac{f^2}{\sqrt{AD}} (dt + c_1 c_5 \omega^0 )^2 + \sqrt{AD} \left[
  \frac{d\rho^2}{\Delta} + d\theta^2 + \frac{\Delta}{f^2} \sin^2 \theta
  d\phi^2 \right].
\end{equation}
We can think of this as a fibration over the three-dimensional base
metric
\begin{equation}
ds_3^2 = \frac{d\rho^2}{\rho^2 - \rho_0^2} + d\theta^2 +
\frac{(\rho^2-\rho_0^2) \sin^2 \theta}{(\rho^2-\rho_0^2) + n^2
  \rho_0^2 \sin^2 \theta} d\phi^2.
\end{equation}
This is exactly the same base metric found in eq. (3.22) of \cite{glr}
(with $(m-n)_{there} = n_{here}$). Thus, passing from the
five-dimensional solutions described there to the four-dimensional one
we consider modifies only the fibration, and not the
three-dimensional base metric, which is what we would expect when adding
a Kaluza-Klein monopole charge. 

As a result, the structure of the four-dimensional metric is the same as in
\cite{glr}. In particular, while the four-dimensional metric is smooth
at $\theta = 0, \pi$, there is a $\mathbb{Z}_n$ orbifold singularity
at $\rho = \rho_0$, and there are curvature singularities in the
three-dimensional base metric at the
corners $\rho=\rho_0$, $\theta=0,\pi$. These curvature singularities
in the base metric do not have simple brane interpretation. Hence, as
in~\cite{glr}, the smooth solutions we have found here do not fit into
the picture of~\cite{bgl}, where supersymmetric solutions were described
as built up from half-BPS atoms.

\section*{Acknowledgments}
We would like to thank Jon Ford, Eric Gimon, Tommy Levi, Samir Mathur
and Yogesh Srivastava for valuable discussions and correspondence. The authors would like to acknowledge the hospitality of the Aspen Center for
Physics during the initial stages of this work. SG
and AS were supported by NSERC. SFR was supported in part by PPARC.

\end{document}